%% file: main.tex
\documentclass{article}

\usepackage{PRIMEarxiv}

\usepackage[utf8]{inputenc} 
\usepackage[T1]{fontenc}    
\usepackage{hyperref}       
\usepackage{url}            
\usepackage{booktabs}       
\usepackage{amsfonts}       
\usepackage{nicefrac}       
\usepackage{microtype}      
\usepackage{lipsum}
\usepackage{graphicx}
\usepackage{titlesec}
\graphicspath{{media/}}     

\usepackage{xcolor}
\usepackage{multirow}
\usepackage{enumitem}
\usepackage{tabularx}

\title{Designing Digital Humans with Ambient Intelligence}

\author{
  Mengyu Chen\thanks{Equal contribution} \quad
  Pranav Deshpande\footnotemark[1] \quad
  Runqing Yang \quad
  Elvir Azanli \quad
  \textbf{Joseph Ligman} \\[0.5em]
  \textbf{Shaohan Hu} \quad
  \textbf{Chun-Fu (Richard) Chen} \\[0.5em]
  JPMorganChase \\[0.3em]
  New York, New York, United States \\[0.3em]
  \texttt{\{mengyu.chen, pranav.deshpande\}@jpmchase.com}
}

\begin{document}

\maketitle

\begin{abstract}
Digital humans are lifelike virtual agents capable of natural conversation and are increasingly deployed in domains like retail and finance. However, most current digital humans operate in isolation from their surroundings and lack contextual awareness beyond the dialogue itself. We address this limitation by integrating ambient intelligence (AmI) - i.e., environmental sensors, IoT data, and contextual modeling - with digital human systems. This integration enables situational awareness of the user’s environment, anticipatory and proactive assistance, seamless cross-device interactions, and personalized long-term user support. We present a conceptual framework defining key roles that AmI can play in shaping digital human behavior, a design space highlighting dimensions such as proactivity levels and privacy strategies, and application-driven patterns with case studies in financial and retail services. We also discuss an architecture for ambient-enabled digital humans and provide guidelines for responsible design regarding privacy and data governance. Together, our work positions ambient intelligent digital humans as a new class of interactive agents powered by AI that respond not only to users’ queries but also to the context and situations in which the interaction occurs.
\end{abstract}


\input{sections/1_introduction}

\input{sections/2_prior_works}
\input{sections/3_concepts}

\input{sections/4_system}
\input{sections/4.5_case_study_implementation}
\input{sections/5_Discussion}

\section*{Disclaimer}
This paper was prepared for informational purposes by the Global Technology Applied Research center of JPMorgan Chase \& Co. This paper is not a product of the Research Department of JPMorgan Chase \& Co. or its affiliates. Neither JPMorgan Chase \& Co. nor any of its affiliates makes any explicit or implied representation or warranty and none of them accept any liability in connection with this paper, including, without limitation, with respect to the completeness, accuracy, or reliability of the information contained herein and the potential legal, compliance, tax, or accounting effects thereof. This document is not intended as investment research or investment advice, or as a recommendation, offer, or solicitation for the purchase or sale of any security, financial instrument, financial product or service, or to be used in any way for evaluating the merits of participating in any transaction.


\bibliographystyle{acm}
\bibliography{references}

\appendix
\newpage
\titleformat{\section}
  {\normalfont\Large\bfseries}{Appendix \Alph{section}:}{0.5em}{}
\input{appendices/appendix_rescoped}

\end{document}

%% file: sections/1_introduction.tex
\section{Introduction}
Digital humans, or lifelike virtual agents capable of natural conversation and expressive behavior, are increasingly used in domains such as retail, health care, travel, and finance \cite{seymour2018actors, bickmore2005social}. These systems now support tasks that range from answering questions in mobile applications to guiding customers through transactions in public service environments. Although recent advances in LLMs have improved the conversation fluency and versatility of these agents \cite{ni2023dialogue, yi2025multiturn}, their awareness of user context and capability to comprehensively understand the user's situation remains extremely limited. Most digital humans rely only on the dialogue channel and execute behaviors only based on user-provided description. As a result, they often fail to adapt to what users are doing, where the interaction is taking place, or which external factors may influence the execution of the task.

Research in ubiquitous computing and ambient intelligence (AmI) demonstrates that environmental sensing, contextual data fusion, and multi-device coordination can significantly enhance the relevance and effectiveness of interactive systems \cite{abowd1999context, baldauf2007context, cook2009ambient}. Modern service environments generate rich streams of contextual information through sensors, enterprise platforms, and personal devices. Prior work shows that these sources can support personalized services, anticipatory assistance, and adaptive system behavior \cite{bettini2010context, gubbi2013iot, pejovic2015anticipatory}. However, the integration of these ideas into digital human technologies is still limited. Most existing systems remain bound to a single device and cannot adjust their behavior based on situational or environmental cues.

We envision the combination of digital humans with AmI creates new opportunities for more adaptive, helpful, and trustworthy assistance to the users. For example, consider a customer entering a bank. Ambient signals such as queue status, appointment information, and human staff availability can help a digital human offer relevant guidance and coordinate with human staff for a smooth check-in experience. Similarly, a user who interacts with their banking application before coming to the bank also generates contextual signals through spending patterns and device activity. Such information can also help a digital human provide timely suggestions and anticipate the needs and intention for their visit. These scenarios illustrate how AmI can shift digital humans \textbf{from reactive conversational interface to proactive agents} that respond to both user actions and environmental context.

However, this direction also raises important challenges. Current digital humans have little access to contextual data, limited continuity across devices, and minimal support for privacy-preserving interpretation of environmental signals. Organizations also face questions about governance, user consent, and accountability when combining sensing infrastructures with autonomous agents \cite{jobin2019ethics}. There is little guidance on how AmI should be structured, designed, or evaluated when used to enhance digital human systems. This gap restricts the deployment of digital humans in domains that require strong contextual awareness, such as finance and health care.

To address this gap, we focus on customer-facing digital humans that operate in both physical service environments, such as retail stores, bank branches, and ubiquitous contexts such as mobile applications and personal wearable devices. Our contributions are:
\begin{itemize} 
    \item A conceptual framework that describe the roles of ambient intelligence in digital human systems, including situational awareness, proactive assistance, cross-device coordination, and personalization.
    \item A design space that identifies key dimensions in the creation of ambient digital humans. These include user constellation, embodiment, levels of proactivity, personalization strategies, temporal scope, and privacy handling.
    \item Application-grounded patterns and architectural considerations, illustrated with case-studies systems from financial and retail services. These examples describe how contextual signals can shape the behavior of a digital human.
    \item An industry-informed analysis of privacy, security, and data governance concerns. We also offer design suggestions for responsible deployment of ambient digital humans.
\end{itemize}
Together, these contributions aim to position ambient digital humans as an emergent class of interactive systems that respond not only to language but also to the situations and environments in which communication takes place.





%% file: sections/2_prior_works.tex
\section{Prior Work}

\subsection{Ambient Intelligence for Contextual Understanding}


Ambient intelligence builds on decades of context-aware computing research, but recent advances in sensing, connectivity, and on-device inference have made it practical to deploy rich contextual systems in real service environments. Comprehensive surveys consolidate the representation and reasoning techniques that translate raw sensor signals into actionable context \cite{bettini2010context}. Sensor-based activity recognition now leverages deep-learning pipelines for robust classification of user actions and intentions \cite{bulling2014activity, lara2013activity, chen2012sensor}. Indoor localization has advanced to BLE, UWB, and visible-light systems with sub-meter accuracy \cite{zafari2019indoor}. IoT architectures provide the connectivity substrate linking distributed sensors at scale \cite{atzori2010iot, gubbi2013iot, stankovic2014iot}. Edge-AI accelerators further make it feasible to run vision and speech models on embedded hardware in service spaces, reducing latency and data-exposure risks \cite{gill2024edge}.

For ambient digital humans, cross-device interaction is especially relevant: users in service environments move fluidly between phones, kiosks, and displays during a single encounter. Brudy et al.\ synthesized over 500 studies into a taxonomy of cross-device interaction patterns \cite{brudy2019crossdevice}, and Houben et al.\ examined practical challenges of such interactions in the wild \cite{houben2017crossdevice}. Most recently, Scargill et al.\ demonstrated how ambient IoT sensors can augment a wearable device's limited perception for richer situational awareness \cite{one_two_three_scargill2023ambientintelligencenextgenerationar}, an approach we extend from AR headsets to embodied conversational agents. Our work integrates multi-layer context (physical sensing, device telemetry, enterprise data) into digital human behavior within the organizational and regulatory constraints of customer-facing service environments that prior AmI research has not systematically addressed.

\subsection{Customer-facing Digital Humans and Virtual Agents}

Interactive virtual agents and digital humans have long been used in customer service, and recent advances are making them more engaging and context-aware. Traditional chatbots have evolved from simple FAQ interfaces to embodied digital humans that exhibit humanlike gestures and empathy in domains such as banking and retail. Relational agents that maintain social rapport across repeated encounters sustain long-term user engagement in public-facing service contexts \cite{bickmore2013tinker}, and dialogue capabilities have been further transformed by large language models, with surveys charting the progression to multi-turn LLM dialogue \cite{ni2023dialogue} and proactive conversational AI, where agents initiate rather than merely respond, emerging as an active frontier \cite{deng2025proactive}. The appearance and persona of a digital human also shape user perception: empirical studies of the uncanny valley caution that near-human but imperfect renderings provoke discomfort \cite{katsyri2015uncanny, macdorman2016uncanny, volonte2016appearance, ho2017uncanny}, while trust frameworks identify dispositional, situational, and learned components of human-automation trust \cite{hoff2015trust}, and anthropomorphism has been shown to increase trust resilience and support longitudinal trust calibration after failures \cite{devisser2016almost, devisser2020longitudinal}. Seymour et al.\ provide a taxonomy of visual presence technologies and their implications for organizational trust \cite{seymour2018actors}.

In customer-facing domains specifically, Xu et al.\ introduced a context-aware 3D virtual agent for financial service that combines mixed reality and vision-language models to enable data-driven, empathetic interactions \cite{four_five_xu2024enablingdatadrivenempatheticinteractions}. Their system integrates situational awareness of the user's physical location (e.g., detecting where the customer is in a bank branch) and personalized assistance based on the customer's profile, while adhering to strict privacy and security requirements \cite{four_five_xu2024enablingdatadrivenempatheticinteractions}. Yuan et al.\ found that customers perceived a celebrity-like digital human agent as more benevolent and trustworthy, and even forgave its mistakes more readily than those of a generic agent \cite{six_hicss56_celebrity_2023}. In clinical settings, virtual agents elicit greater self-disclosure than human interviewers, particularly in mental-health screening and veteran outreach \cite{devault2014simsensei, lucas2014computer, philip2017virtual, rizzo2019virtual}. These studies underscore the importance of context and design in customer-facing agents: by incorporating environmental context and social cues, digital humans can provide more relevant help and build rapport. Our work extends this trend by combining the embodied, relational qualities of digital humans with ambient sensing and enterprise data integration-adding environmental awareness, cross-device continuity, and context-driven proactivity that most prior systems lack.

\subsection{Agentic Tool Use and Multi-Modal Interfaces}

Beyond conversational prowess, modern AI agents are increasingly \emph{agentic} able to use external tools, APIs, and multi-modal inputs to achieve goals autonomously. Toolformer and Gorilla demonstrated that language models can learn to invoke APIs on their own \cite{schick2023toolformer, patil2023gorilla}, and broader surveys map the rapidly expanding tool-learning landscape \cite{qin2023tool}. These capabilities rest on structured reasoning techniques such as chain-of-thought prompting for multi-step inference \cite{wei2022chain} and ReAct for interleaved reasoning and action \cite{yao2023react}, and are complemented by retrieval-augmented generation that grounds model outputs in retrieved evidence \cite{lewis2020rag, gao2024rag}. A number of frameworks now allow multiple specialized AI agents to be orchestrated together. Microsoft's AutoGen framework allows developers to create applications where multiple LLM-driven agents converse with each other and call tools as needed, flexibly incorporating human input into the loop \cite{nine_sixteen_seventeen_wu2023autogenenablingnextgenllm}. MetaGPT assigns agents distinct software-engineering roles governed by standard operating procedures \cite{hong2024metagpt}, and Park et al.'s generative agents simulate believable social behavior through memory streams and reflection, revealing emergent group dynamics \cite{park2023generative}. The OpenHands platform extends these concepts by enabling AI agents that can write code, run command-line operations, and browse the web autonomously within a sandboxed environment that supports multi-agent coordination \cite{eight_wang2025openhandsopenplatformai}. Comprehensive surveys map the taxonomy of LLM-based agents \cite{wang2023agents, xi2023agents, guo2024multiagents}, and open interoperability standards (e.g., Model Context Protocol \cite{anthropic2024mcp} and Agent-to-Agent protocol \cite{google2025a2a}) are beginning to connect agents with data sources, tools, and one another through standardized and other agents respectively.

The state of the art is moving toward multi-modal, tool-augmented agents and AI systems that do not just chat, but can see through vision models, act through code and tools, and collaborate with other agents. Multi-modal foundation models such as GPT-4 and Gemini have expanded what agents can perceive by processing interleaved text, image, and audio \cite{openai2023gpt4, geminiteam2023gemini, yin2024multimodal}. Computer use agents operating GUIs and browsers represent a frontier with clear limitations where benchmarks such as WebArena, Mind2Web, and OSWorld report that frontier models achieve only 12--22\% on realistic desktop tasks \cite{zhou2024webarena, deng2023mind2web, xie2024osworld}, and commercial deployments confirm that failure modes persist in production \cite{anthropic2024computeruse, openai2025operator}. The question of when agents should act autonomously connects to mixed-initiative interaction research: Amershi et al.\ distilled 18 empirically grounded guidelines for human-AI interaction \cite{amershi2019guidelines} that inform our system's graduated initiative levels, human-handoff protocols, and transparency mechanisms. These capabilities inform our system design and we equip the digital human with modules for perception and action, not just conversation, synthesizing tool-augmented reasoning, multi-agent orchestration, and multi-modal perception within a unified architecture tailored to the constraints of regulated, customer-facing service environments.

%% file: sections/3_concepts.tex
\section{Ambient Intelligence Driven Digital Humans Concept}
Ambient intelligence can expand digital human capabilities by providing situational awareness and environmental affordances that are not accessible through conversation alone. Prior works on digital humans, embodied agents, and conversational systems rarely examines how environmental context, organizational processes, or cross-device signals can shape interaction. Research in ubiquitous computing and context-aware systems shows that contextual information can improve task efficiency, reduce cognitive load, and support more adaptive behavior. However, the application of these ideas to digital human experiences remains limited and largely fragmented. 

To address this gap, we build on established principles in context modeling \cite{gay2004activity}, multi-device interaction \cite{one_two_three_scargill2023ambientintelligencenextgenerationar}, and service system design \cite{carrerarivera2022context}. From these foundations, we derive a conceptual framework for understanding how AmI can support digital humans. The framework is structured around two questions:
\begin{enumerate}
    \item What roles can ambient intelligence play in shaping digital human behavior?
    \item what kinds of contextual layers provide the information and action channels that digital humans can draw upon?
\end{enumerate}
The remainder of this section answers these two questions by identifying the key roles of AmI in digital human experiences and by describing the ambient context layers that support them. We then present a design space that characterizes how ambient digital humans can vary across application settings.

\subsection{Roles of Ambient Intelligence in Digital Human Experiences}
The idea of characterizing digital human capabilities through a set of roles draws from prior work that defines functional axes for context-aware systems, including situational, social, temporal and environmental context \cite{dey2001framework, schmidt1999context, erickson2000social}. Similar work in embodied agents and service robotics highlights the importance of accessing environmental state, organizational constraints, and predictive cues \cite{wirtz2018service}. We identify five roles that AmI can play in shaping digital human experiences.

\textbf{\textit{R1: Situational Awareness}} - 
AmI can inform the digital human about the location and setting of the interaction, the presence of other individuals, and the activities in the space using a variety of sensors and data sources. Situational awareness is fundamental in context-aware computing and has been shown to improve the relevance and efficiency of system responses \cite{dey2001understanding}. For digital humans, this means avoiding unnecessary questions, presenting shortcuts to support materials, and aligning with what the user is currently experiencing.

\textbf{\textit{R2: Proactive and Anticipatory Interaction}} -
Anticipatory computing research demonstrates that the system can act more helpfully when they respond to predicted needs or environmental cues \cite{pejovic2015anticipatory}. AmI offers digital humans access to signals that reveal upcoming events or possible user intentions, allowing the agent to act proactively in a way that aligns with user expectations without explicit user prompts.

\textbf{\textit{R3: Multi-User Conversation and Social Context Handling}} -
Ami empowers digital humans to comprehend and participate in fluid, multi-user conversations, moving beyond traditional turn-based question-answer interactions. By continuously monitoring the social context, the digital human can track multiple speakers, interpret overlapping dialogues, and identify when its input is needed, such as clarifying misunderstandings, mediating group decisions, or offering timely suggestions. For example, in a family banking scenario, the agent can seamlessly join a discussion about shared accounts, ensuring privacy and consent for each participant, while in a retail setting, it can recognize group dynamics and provide tailored recommendations or assistance to individuals or the group as a whole.

\textbf{\textit{R4: Adaptive Modality and Cross-Channel Presence}} -
Multi-device and cross-surface interaction research highlights how users fluidly move between phones, kiosks, displays, and other devices during a task \cite{brudy2019crossdevice}.
With AmI, digital humans can dynamically adjust their mode of interaction and level of presence based on situational factors. They can recognize the interface transitions and coordinate actions across channels. For example, in an airport setting, the agent might switch between user's phone and kiosk to provide clear on-screen instructions.

\textbf{\textit{R5: Continuous Learning and Personalization}} -
Personalization research shows that long-term information aout user preferences and prior interactions can enhance trust, recommendation, and support more efficient task completion \cite{fischer2001user}. AmI supports ongoing learning from both user interactions and environmental feedback, enabling digital humans to refine their understanding of user preferences, routines, and needs over time. This continual adaptation could lead to highly personalized experiences, such as customized financial advice that evolves with the user’s life events or tailored health recommendations based on daily activity patterns.

\subsection{Ambient Context Layers for Digital Humans}
To support the roles described above, AmI for digital humans needs to draw from multiple layers of contextual information and to be able to act through different channels. The concept of ambient context layers is inspired by prior models that describe context as a structured combination of physical, device-level, and spatial information \cite{perera2014context}. We adopt this layered perspective to organize how digital humans access, memorize, and act on context. \autoref{tab:designspace} shows a list of ambient context layers and their examples, application domains, and their relations with the AmI roles.

\subsubsection{Layers for Contextual Information Retrieval}
To give digital humans a clear understanding of their surroundings and the situations they operate in, we envision four layers of contextual information. Each layer captures a different aspect of the environment or the user's activity that can help the agent interpret what is happening:

\textbf{\textit{Physical sensing layer}} - Environmental sensors such as cameras, microphones, motion detectors, and beacons reveal the physical conditions of the interaction space. Such signals could support location inference, activity recognition, social presence detection, and environmental awareness.

\textbf{\textit{Device and application layer}} - Signals from mobile applications, kiosks, displays, and terminals reveal ongoing tasks, device capabilities, and user activity states. This layer supports information synchronization and coordination across multiple surfaces.

\textbf{\textit{Enterprise infrastructural layer}} - Information from CRM platforms, policy engines, transaction logs, and risk models provides institutional knowledge that influences what actions are possible or appropriate. This layer is necessary for tasks that involve compliance, authorization, or operational rules.

\subsubsection{Layers for System Actuation}
For digital humans to respond effectively, they must be able to act through channels that communicate with the user or influence the environment. We envision three layers that enable such action:

\textbf{\textit{Environmental layer}} - Digital signage, displays, lighting cues, and audio outputs enable the digital human to guide the user's attention or shape the experience in the surrounding environment by modifying the artifact properties.

\textbf{\textit{Conversational layer}} - Speech, text, on-screen avatars, or other representations allow the digital human to render itself and communicate information with the user.

\textbf{\textit{Utility action layer}} - This includes account operation, workflow triggers, document generation, and other back-end transactional behaviors that allow the digital human to complete tasks on the user's behalf.

\begin{table}[t]

\centering

\begin{tabularx}{\linewidth}{l X l X}

\toprule

\textbf{Layer} & \textbf{Examples} & \textbf{Supported Roles} & \textbf{Domains} \\

\midrule


\multicolumn{4}{l}{\textbf{Information Retrieval}} \\

\cmidrule(lr){1-4}

Physical sensing & Cameras, microphones, proximity sensors & R1, R3 & Banking, retail, hospitality \\

Cross-device interaction & Mobile app state, kiosk events & R1, R4 & Public services, retail, banking \\

Infrastructural platforms & CRM, policy engines, risk systems & R2, R3, R5 & Banking, healthcare, public services \\


\multicolumn{4}{l}{\textbf{System Actuation}} \\

\cmidrule(lr){1-4}

Conversational & Speech, text, human-like avatar & R4, R5 & All domains \\

Environmental & Displays, signage, lighting cues & R1, R4 & Retail, hospitality \\

Utility action & Workflows, documents, transactions & R2, R3, R4 & Banking, healthcare, public services \\

\bottomrule

\end{tabularx}

\caption{Ambient context layers for digital humans, grouped by information retrieval and agentic action.}

\label{tab:contextlayers}

\end{table}

\subsection{Design Space}
\label{sec:designspace}

The roles (R1–R5) describe \emph{what} ambient intelligence enables for digital humans; the ambient context layers specify \emph{where} the supporting signals originate and \emph{through which channels} actions are delivered. A complementary question remains: \emph{how do concrete ambient digital human deployments differ from one another, and from conventional conversational agents?} To answer this, we introduce a two-dimensional design space defined by two fundamental axes: 1) \textbf{context richness}, and 2) \textbf{system initiative}.

The rationale for organizing the design space along these two axes draws from established taxonomies in context-aware computing, mixed-initiative interaction, and multi-device design \cite{horvitz1999mixed, brudy2019crossdevice}. Prior work consistently identifies two overarching sources of variation in intelligent interactive systems: the breadth and depth of environmental information the system can access, and the degree of autonomy with which it acts on that information. By casting these as orthogonal dimensions, the design space provides a compact frame for comparing system configurations, identifying gaps, and reasoning about trade-offs. \autoref{tab:quadrants} characterizes the four resulting quadrants as system archetypes.

\subsubsection{Axis 1: Context Richness}

Context richness captures \emph{how much the system knows} about the user, the environment, and the organizational setting at the moment of interaction. It spans a continuum from dialogue-only systems that rely exclusively on the user's spoken or typed input, to fully ambient systems that fuse signals from the physical sensing, device and application, and enterprise infrastructural layers described in the previous subsection.

Several design sub-dimensions determine a system's position along this axis:

\textbf{Observability across context layers.} Which signals are available, from the \textit{Physical sensing} layer (e.g., entry and occupancy indicators, ambient sound levels), the \textit{Device and application} layer (e.g., mobile session state, kiosk events), and the \textit{Enterprise infrastructural} layer (e.g., CRM flags, eligibility rules, policy constraints). Designs can range from minimal sensing in privacy-first deployments to rich multi-layer observability in operational contexts such as bank branches or hospitals.

\textbf{Situational state and granularity.} Which state variables are inferred (e.g., user intent hypotheses, environment state, audience topology, device availability, time pressure) and at what temporal resolution (per-event, per-session, rolling window). Finer granularity improves responsiveness and cross-device continuity; coarser granularity reduces processing and data retention burdens.

\textbf{Uncertainty and calibration.} How confidence in inferred states is represented and used. Calibrated uncertainty estimates determine whether the system should remain silent, hint, suggest, or escalate, and provide a mechanism for graceful degradation when signals are noisy or conflicting.

\textbf{Data quality, alignment, and provenance.} How inputs are validated and reconciled across layers: device localization and time alignment, schema validation, deduplication, and provenance tags indicating source and signal quality. Provenance-aware fusion improves robustness and informs appropriate disclosure and action choices, particularly in multi-user settings.

\subsubsection{Axis 2: System Initiative}

System initiative captures \emph{how autonomously the system acts}, the degree to which the digital human can anticipate needs, choose actions, and execute operations without waiting for explicit user prompts. The continuum ranges from purely reactive systems that respond only to direct user requests, to proactive agents that monitor contextual signals, predict needs, and act preemptively within authorized boundaries.

Several design sub-dimensions determine a system's position along this axis:

\textbf{Initiative and timing.} When the agent may act unprompted and through which channels. Initiative is modulated by situational confidence and user consent and follows a graduated scale: silent, hint, suggest, prefill, or act. The graduation ensures helpfulness scales with evidence rather than being driven by fixed rules.

\textbf{Cross-channel continuity and selective disclosure.} How task state and identity move across the \textit{Conversational} and \textit{Environmental} channels, while ensuring that personally identifiable information never appears on shared surfaces. Continuity tokens or equivalent constructs enable private handoff from public to personal devices, while personalization remains scoped and revocable.

\textbf{Actuation scope and safeguards.} What the agent is authorized to do via the \textit{Utility action} layer, from low-risk guidance and wayfinding, to medium-risk prefill and document generation, up to high-risk transactional operations. Each scope tier binds to appropriate consent and confidence requirements and may require dual control with human staff for higher-risk actions.

\textbf{Human handoff and organizational coupling.} How and when human staff are notified, how organizational roles and availability affect routing, and how accountability is recorded. This sub-dimension ensures agent behaviors align with institutional processes, especially under exceptions and edge cases.

\subsubsection{System Archetypes}

The intersection of context richness and system initiative produces four quadrants, each representing a recognizable system archetype (\autoref{tab:quadrants}). These archetypes are not rigid categories but regions of the continuous space, useful for characterizing existing systems and identifying the design opportunity that motivates this work.

\begin{table}[t]
\centering
\renewcommand{\arraystretch}{1.35}
\begin{tabularx}{\linewidth}{>{\bfseries}l X X}
\toprule
 & \textbf{Low Context Richness} & \textbf{High Context Richness} \\
\midrule
\rotatebox[origin=c]{0}{\parbox{2.0cm}{\bfseries High\\Initiative}} &
\textit{Eager Uninformed Agent.}\newline Acts proactively but lacks environmental grounding. May volunteer suggestions that are irrelevant or poorly timed, increasing the risk of user annoyance or trust erosion. &
\textit{Ambient Digital Human.}\newline Fuses multi-layer context with proactive, safeguarded action. Anticipates needs, coordinates across channels, and acts within governed boundaries, the target archetype of this work. \\
\midrule
\rotatebox[origin=c]{0}{\parbox{2.0cm}{\bfseries Low\\Initiative}} &
\textit{Conventional Conversational Agent.}\newline Responds to explicit user prompts using dialogue history alone. Effective for simple question-answering but unable to adapt to situational cues or coordinate across devices. &
\textit{Context-Aware Assistant.}\newline Observes rich environmental and enterprise signals but waits for user requests before acting. Reduces redundant questions and personalizes responses, yet misses opportunities for anticipatory support. \\
\bottomrule
\end{tabularx}
\caption{Four system archetypes arising from the design space. Each quadrant describes a characteristic combination of context richness and system initiative, with the ambient digital human (top-right) representing the vision advanced in this paper.}
\label{tab:quadrants}
\end{table}

Most existing digital humans and virtual assistants operate in the lower-left quadrant: they are dialogue-bound and reactive. Some systems have begun to incorporate enterprise data or device signals, moving them rightward into the \textit{Context-Aware Assistant} region, while others leverage large language models to offer unsolicited suggestions, edging upward into the \textit{Eager Uninformed Agent} region without sufficient grounding. The ambient digital human archetype, situated in the upper-right quadrant, combines deep contextual awareness with calibrated proactive behavior, and it is this combination that necessitates the governance constraints described next.

%% file: sections/4_system.tex
\section{Ambient Sensing and Context Inference}
\label{sec:sensing}
The conceptual framework presented in the previous section identified five roles for ambient intelligence (R1--R5) and organized the supporting infrastructure into layers for contextual information retrieval and system actuaion. This section describes how the information retrieval layers from physical sensing, device and application signals, to enterprise infrastructure are realized in practice. We ground each mechanism in the design space axes introduced earlier, focusing on: 1) observability across context layers, 2) situational state and granularity, and 3) data quality and provenance. The subsequent section then addresses the action layers.

We focus on customer-facing service environments, specifically, bank branches and retail locations, where the digital human operates as a situated agent embedded in a physical space with access to multiple sensing modalities and enterprise data sources. We chose these specific domains because they exemplify the challenges and opportunities of ambient digital humans: they involve complex service workflows, require strong contextual awareness for effective human-digital communication, and are subject to stringent privacy control and regulatory constraints. In these settings, contextual signals arise from three broad sources: (1) the physical environment, sensed through microphones, cameras, and spatial infrastructure; (2) the digital devices that users carry or interact with, including smartphones, kiosks, and terminals; and (3) the institutional systems that govern service delivery, such as appointment schedulers, customer relationship management (CRM) platforms, and risk engines. Each source contributes a different facet of the situational state that the digital human uses to calibrate its behavior.

\subsection{Physical Sensing Network}

The physical sensing layer captures real-time signals from the interaction space through microphones, cameras, motion detectors, and spatial infrastructure such as beacons and occupancy sensors. In our implementation, the physical sensing network is structured around three modalities: voice input and paralinguistic cue detection, visual presence sensing and identification, and environmental and spatial understanding. Each modality contributes distinct state variables to the agent's situational model (R1) and provides triggers for anticipatory behavior (R2).

\begin{figure}[t]
  \centering
  \includegraphics[width=\linewidth]{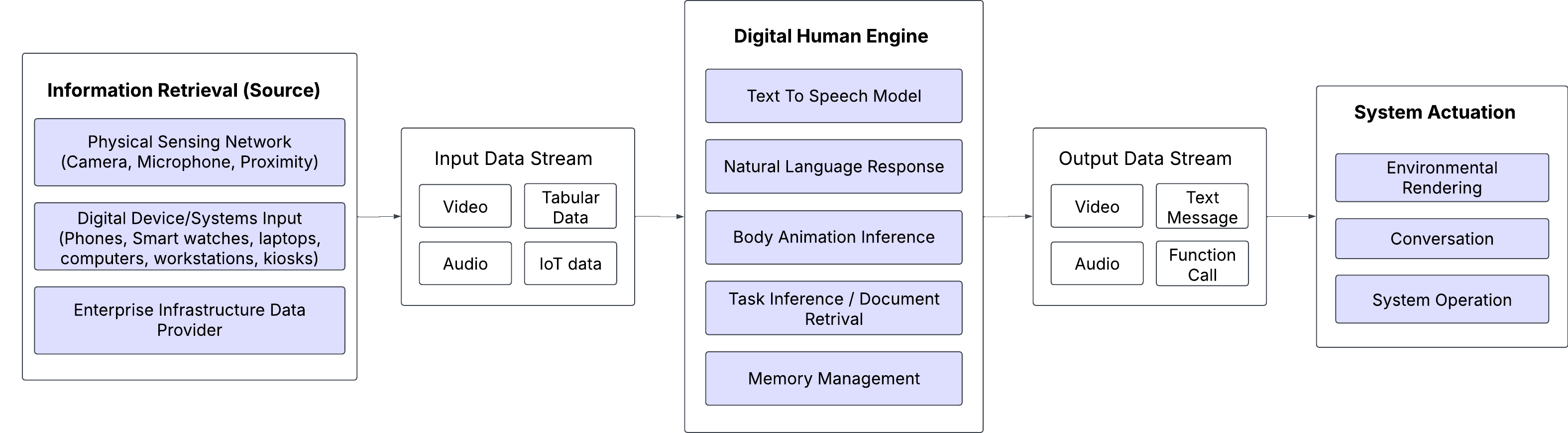}
  \caption{A conceptual illustration of an ambient intelligence-enhanced digital human framework. The digital human engine(center) receives multi-layered contextual inputs – physical environment, device activity, enterprise data, social cues – and plays roles such as situational awareness and proactive assistance (R1–R5 from Section 3.1). This ambient context enables the agent to move from a purely reactive chatbot to a proactive assistant attuned to the user’s context.}
  \label{fig:diagram1}
\end{figure}

\subsubsection{Voice Input and Cue Detection}

Voice input serves as the primary conversational channel between the user and the digital human, but it also carries ambient signals that extend beyond the spoken content. The system captures audio through microphones positioned in the service environment (e.g., embedded in a kiosk or mounted at a service counter) and processes it through a dedicated audio pipeline. Speech-to-text transcription is performed by a transformer-based ASR model, in our implementation, OpenAI Whisper \cite{radford2023whisper}, which supports multilingual transcription and produces streaming partial results with sub-second latency. Alternative real-time ASR options include Google Cloud Speech-to-Text and NVIDIA Riva, each offering trade-offs between on-device privacy and cloud-based accuracy \cite{prabhavalkar2024asr}. A voice activity detector (VAD) segments the audio stream into utterance boundaries, identifying pauses and turn completions to support natural turn-taking \cite{skantze2021turntaking}. Our implementation uses a WebRTC-based VAD with energy-threshold gating; lightweight neural alternatives such as Silero VAD \cite{silero2021vad} can further improve robustness in noisy environments at minimal computational overhead ($<$1\,ms per frame on CPU).

Beyond transcription, the audio pipeline extracts paralinguistic cues that inform the agent's situational awareness (R1). Prolonged silence, hesitation markers (e.g., filled pauses, restarts), and shifts in vocal tone can signal user uncertainty, frustration, or confusion. The system monitors these cues through event-driven triggers: for instance, a silence exceeding a configurable threshold (e.g., $>$2\,s in our implementation) after a system prompt may cause the agent to offer clarifying assistance (e.g., \textit{``It seems like you might have a question, would you like me to help?''}). This mechanism supports proactive and anticipatory interaction (R2) by allowing the digital human to respond to behavioral signals rather than waiting for explicit verbal requests. The design draws on research in conversational grounding and repair in spoken dialogue systems \cite{clark1991grounding}, which demonstrates that attending to paralinguistic features can reduce communication breakdowns and improve task success rates.

In multi-user scenarios (R3), the audio pipeline must also distinguish between speakers. The current implementation supports single-user voice interaction, but the architecture accommodates neural speaker diarization models such as pyannote.audio \cite{bredin2023pyannote} that can attribute utterances to different individuals in near real time. This is used in settings where multiple customers or staff members may be present near the agent simultaneously.

\subsubsection{Visual Presence Sensing and Identification}

The visual sensing modality provides the digital human with awareness of user presence, position, and activity state in the physical space. Cameras positioned at service points or integrated into kiosk hardware capture video streams that are processed through real-time object detection models, such as YOLOv8 \cite{jocher2023yolov8} or lightweight alternatives (e.g., MobileNet-SSD) suitable for edge deployment, to detect and track individuals within the interaction zone. Detected bounding boxes are fed to a multi-object tracker (e.g., ByteTrack \cite{zhang2022bytetrack} that maintains consistent identities across frames. Visual processing supports two primary functions: presence detection and activity state inference.

Presence detection determines whether a user is within the agent's interaction range and estimates their spatial position relative to the service point. This information enables the digital human to initiate engagement at contextually appropriate moments, for example, greeting a user who approaches the kiosk rather than requiring them to press a button or speak first. Prior work on proxemic interaction has shown that spatial awareness enables more natural and less intrusive system initiation \cite{ballendat2010proxemic}. In our system, presence signals also contribute to queue awareness: by tracking the number and positions of individuals in a service area, the agent can estimate wait times and adjust its behavior accordingly (e.g., prioritizing self-service options during peak periods).

Activity state inference operates at a higher level, using visual cues to estimate what the user is currently doing, whether they are actively engaged with the kiosk, looking at their phone, speaking with another person, or appearing to wait. These state estimates feed into the agent's situational model (R1) and modulate its initiative level: the agent may suppress proactive prompts when the user appears occupied with another task and re-engage when attention returns. In the current implementation, activity classification relies on lightweight pose and gaze estimation models such as MediaPipe \cite{lugaresi2019mediapipe} for body-pose and head-orientation estimation, and gaze-tracking modules for attention inference. Importantly, the system does not perform facial recognition for identification purposes; instead, identity is established through explicit authentication mechanisms (e.g., scanning a QR code from a mobile device, tapping an NFC-enabled card, or entering credentials at a terminal), preserving user privacy while still enabling personalized interaction once identity is confirmed through a consented channel.

\subsubsection{Environmental and Spatial Understanding}

Beyond sensing individual users, the physical sensing layer captures broader environmental conditions that shape the service context. Ambient signals such as noise levels, lighting conditions, occupancy counts, derived from overhead depth cameras (e.g., Intel RealSense) or time-of-flight people counters deployed at entry points, and time of day provide the digital human with a coarse-grained understanding of the overall state of the space. These signals support the environmental and spatial awareness needed for the agent to calibrate its behavior to the setting rather than only to the individual user.

In a bank branch, for example, occupancy and queue data inform the agent about the current service load. If the branch is crowded, the digital human may prioritize concise guidance and direct users toward self-service options; during quieter periods, it may offer more extended, exploratory assistance. Queue status is inferred from a combination of ticket system data and spatial occupancy signals, and is surfaced to the agent as a contextual variable that modulates both response content and initiative timing (R1, R2). Similarly, appointment and scheduling data from the branch management system can be cross-referenced with physical presence signals to identify when a user with a known appointment has arrived, enabling the agent to proactively acknowledge them and streamline check-in.

Spatial infrastructure such as Bluetooth Low Energy (BLE) beacons, Ultra-Wideband (UWB) anchors, or Wi-Fi Round-Trip Time (RTT) positioning \cite{zafari2019indoor} provides indoor localization at varying levels of accuracy (1--3\,m for BLE, $<$30\,cm for UWB), helping the agent determine whether a user is near the entrance, at a service counter, or in a waiting area. This positional context supports adaptive modality selection (R4) by informing which output channel is most appropriate,a nearby wall display, the user's personal device, or the kiosk speaker. Environmental signals are collected through context adapters that poll or subscribe to sensor data streams (typically via MQTT or REST endpoints) and normalize the readings into a shared JSON representation consumed by the ambient intelligence engine. This normalized context state is continuously updated and made available to downstream reasoning processes, ensuring that the agent's situational model reflects the current conditions of the service environment.

\subsection{Digital Device and System Inputs}

The device and application layer captures signals from the digital surfaces and personal devices that users interact with during a service encounter. In customer-facing environments, these surfaces typically include self-service kiosks, in-branch terminals, mobile banking applications, and web portals. Each device generates contextual signals, such as active application state, navigation history, session metadata, and authentication events, that the digital human can draw upon to maintain continuity and calibrate its assistance.

Personal devices play a particularly important role in bridging physical and digital contexts. A user's smartphone, for example, may carry session tokens (e.g., a JWT or OAuth~2.0 access token) from a prior interaction with a mobile banking application. When the user arrives at a branch and authenticates at a kiosk (e.g., by scanning a QR code displayed on their phone), the system can retrieve that session context via a secure API call and continue the task seamlessly. In our prototype, this handoff is mediated by a WebSocket connection between the mobile client and the kiosk front-end, with session state serialized as a signed JSON payload. This mechanism supports cross-channel continuity (R4) by enabling the digital human to resume an interaction that began on a different surface, for instance, recognizing that a user who started a loan application on their phone may now wish to complete it in person with additional guidance. Prior work on multi-device interaction demonstrates that such handoff capabilities reduce the cognitive overhead of context-switching and improve task completion rates \cite{brudy2019crossdevice, houben2017crossdevice}.

Beyond session continuity, device-level signals also inform the agent about which interaction modalities are available at a given moment. If a user is interacting at a kiosk equipped with a screen and speaker, the agent can present visual information alongside spoken dialogue. If the interaction shifts to a mobile device, the agent may adapt its output to a smaller display or rely more heavily on text-based communication. This modality negotiation aligns with the design space axis of \emph{cross-channel continuity and selective disclosure}: the system selects the appropriate rendering channel based on device capabilities, environmental context, and privacy constraints. Notably, the system ensures that personally identifiable information is never displayed on shared surfaces; instead, sensitive content is routed to the user's personal device, where it can be accessed privately.

Device-level signals also contribute to situational awareness (R1). The system can detect whether a user's mobile application is currently active, which screen or workflow the user is viewing, and whether a prior session was abandoned or completed. These signals help the digital human avoid redundant prompts, for instance, skipping an introductory question if the user has already provided relevant details through the mobile application. In aggregate, the device and application layer extends the agent's observability beyond the physical space and into the digital interactions that precede, accompany, or follow a service encounter.

\subsection{Enterprise Infrastructure Data Providers}

The enterprise infrastructural layer connects the digital human to the institutional knowledge systems that govern service delivery. Unlike physical and device-level signals, which capture the immediate state of the interaction environment, enterprise data provides the organizational, regulatory, and historical context that determines what actions are permissible, what information is relevant, and how the agent should frame its assistance. In customer-facing domains such as banking and retail, this layer encompasses CRM platforms, appointment and queue management systems, transaction ledgers, and risk and policy engines \cite{maglio2009service}. The digital human accesses these systems through thin API connectors (REST or GraphQL) that query enterprise endpoints and normalize the returned data into a shared JSON context representation used by the ambient intelligence engine. We organize enterprise data provision around three functional roles: session data retrieval, historical data retrieval, and risk and policy gating.

\subsubsection{Session Data Retrieval}

Session data encompasses the transient, visit-scoped information that characterizes a user's current service encounter. This includes appointment records, queue registrations, check-in status, and any service requests initiated during the visit. When a user authenticates at a kiosk or is identified through a device-level handoff, the system queries the branch management platform via a RESTful API call (typically keyed on user ID or appointment reference) to retrieve their active session context. For instance, if the user has a scheduled appointment for a mortgage consultation, the digital human can immediately acknowledge this purpose, direct the user to the appropriate waiting area, and notify the relevant staff member of the user's arrival, initiating behaviors that exemplify proactive and anticipatory interaction (R2).

Session data also supports the agent's understanding of the user's position within a service workflow. If the user has already completed certain steps (e.g., identity verification at a prior touchpoint), the agent can skip redundant prompts and advance the conversation to the next relevant stage. This capability reduces interaction friction and aligns with the design space axis of \emph{situational state and granularity}: the agent maintains a fine-grained, per-event model of the user's progress through the service process. Importantly, session data is scoped to the current visit and does not require access to sensitive historical records, making it a lower-risk entry point for contextual enrichment.

\subsubsection{Historical Data Retrieval}

Historical data extends the agent's awareness beyond the current session to encompass prior interactions, past transactions, and longitudinal user attributes. This information is drawn from CRM systems, transaction ledgers, and interaction logs maintained by the organization. Access to historical data enables the digital human to personalize its assistance (R5), for example, referencing a loan inquiry from a previous visit, acknowledging a recent account change, or adapting its communication style based on recorded user preferences.

However, historical data is inherently more sensitive than session-scoped information. Transaction histories, account balances, and prior service records constitute personally identifiable information (PII) subject to regulatory requirements such as GDPR and financial sector compliance standards (e.g., KYC). The system therefore treats historical data retrieval as a gated operation: access is mediated by the risk and policy layer described below, and the scope of retrieved data is limited to what is relevant to the current interaction context. The design space axis of \emph{consent choreography and purpose limitation} governs this boundary, and the agent retrieves historical information only when a legitimate purpose exists and the user has provided appropriate consent, either explicitly during the session or through prior opt-in at account enrollment. When historical data is used to inform the agent's behavior, the system can disclose this to the user (e.g., \textit{``Based on your recent activity, it looks like you may be interested in our savings options''}), supporting the transparency goals described in the design space.

\subsubsection{Risk and Policy Engines}

Risk and policy engines serve as the gating component that governs what contextual information the digital human may access and what actions it may initiate. These engines encode institutional rules, regulatory constraints, and risk thresholds that modulate the agent's behavior in accordance with organizational policies. In our design, every data retrieval request and proposed action is evaluated against a policy layer before execution.

For data access, the policy engine determines the scope and granularity of information the agent can retrieve based on the current use case, the user's consent status, and the sensitivity classification of the requested data (mapped to classes such as public, internal, confidential, and restricted). A routine queue inquiry, for instance, requires no special authorization, while retrieving detailed transaction history may require a higher level of authentication or an active consent token. This design aligns with the \emph{data quality, alignment, and provenance} axis of the design space: each data element carries provenance metadata indicating its source, sensitivity level, and the conditions under which it was obtained.

For action gating, risk engines evaluate proposed operations against fraud detection models and compliance rules in real time. If the agent detects that a user is about to authorize an unusual transaction, or if conversational cues suggest a potential social engineering scenario, the risk engine can flag the interaction, prompting the digital human to intervene with a verification prompt or escalate to a human staff member \cite{ali2022fraud}. This mechanism supports anticipatory assistance (R2) in the domain of user safety: the agent acts not only to fulfill requests but also to protect the user from potentially harmful outcomes. The risk and policy layer thus functions as both a safeguard and an enabler, ensuring that the digital human operates within institutional boundaries while still leveraging enterprise data to provide contextually informed assistance.

\section{System Actuation and Orchestrated Behaviors}
\label{sec:actuation}

Where the preceding section described how the digital human gathers and interprets contextual signals, this section addresses how it acts by rendering assistance, communicating with users, and executing operations on their behalf. These capabilities map to the three action layers defined in the conceptual framework: the environmental layer, through which the agent shapes the physical surroundings; the conversational layer, through which it communicates via speech, text, and embodied representation; and the utility action layer, through which it performs transactional and workflow operations. The design space axes governing these actions: \emph{initiative and timing}, \emph{cross-channel continuity and selective disclosure}, \emph{actuation scope and safeguards}, and \emph{human handoff and organizational coupling}.

A key design principle across all three action layers is that the digital human's behavior is modulated by the situational state assembled from the sensing layers. Rather than following fixed scripts, the agent selects its actions based on the current context: the user's position in a service workflow, the available interaction surfaces, the sensitivity of the information involved, and the confidence of its situational inferences. This context-driven orchestration is what enables the shift from reactive conversational interface to proactive ambient agent described in the introduction.

\subsection{Environmental Rendering}

The environmental layer enables the digital human to communicate through the physical space itself, using shared displays, digital signage, lighting cues, and audio outputs to guide user attention, convey non-sensitive information, and coordinate spatial navigation. This layer extends the agent's presence beyond the conversational interface and into the broader service environment \cite{vogel2004ambient}.

In our design, environmental rendering serves two primary functions. First, it provides ambient guidance that does not require direct conversational engagement. Shared displays in a branch lobby, for example, can show general queue status, estimated wait times, or wayfinding instructions that help users orient themselves upon entry. These outputs are driven by the agent's situational model and updated in real time via MQTT or WebSocket push from the ambient intelligence engine in response to changes in occupancy, queue state, or service availability. Because shared displays are visible to all individuals in the space, the system enforces a strict constraint: no personally identifiable information is rendered on environmental surfaces. This \emph{selective disclosure} principle, articulated in the design space, ensures that PII is routed exclusively to personal devices or authenticated terminals.

Second, environmental rendering supports private handoff between shared and personal channels. When the digital human needs to convey sensitive content, such as account details or a personalized recommendation, it can display a QR code on a public screen or send a deep link to the user's mobile device, prompting the user to continue the interaction on a private surface. This mechanism preserves the fluidity of the interaction (R4) while respecting privacy boundaries. The handoff is mediated by a cryptographically signed continuity token (JWT) that links the public-surface session to the user's authenticated identity on their personal device, enabling the conversation and task state to transfer seamlessly without requiring the user to re-enter information.

In the current prototype, environmental outputs are simulated through a web-based dashboard that renders the signage and display content the system would produce in a deployed setting. The orchestration layer, however, is designed to drive physical signage and display hardware through a standardized REST interface (accepting structured display payloads) when connected, making the architecture extensible to production environments. The inclusion of environmental rendering as a first-class action layer reflects the AmI principle that the agent should operate \emph{through} the environment, not merely \emph{within} a single conversational window, allowing it to shape the user's experience across the full spatial context of the service encounter.

\subsection{Conversations}

The conversational layer is the primary channel through which the digital human communicates with users. It encompasses spoken dialogue, text-based interaction, and the on-screen embodied representation of the agent. Unlike traditional conversational systems that generate responses based solely on dialogue history, the ambient digital human's conversational behavior is continuously shaped by the contextual state assembled from the sensing layers described in the preceding section. This context-enriched approach to conversation is central to realizing situational awareness (R1), proactive assistance (R2), and personalization (R5).

\subsubsection{Context-Enriched Dialogue Generation}

At the core of the conversational layer is a language model that generates the agent's verbal and textual responses. The system employs a large language model (LLM) as the generative backbone, served through Ollama \cite{ollama2024} running a locally deployed model (e.g., Llama~3 or Mistral), which provides full data residency, an important consideration in regulated domains. The ambient intelligence engine, implemented as a Flask microservice, queries the model server at each conversational turn. Rather than passing only the dialogue history to the model, the engine composes a context-augmented prompt that integrates the current conversational context with relevant situational signals: the user's position in the service workflow, applicable enterprise data (e.g., appointment details, account flags), environmental conditions (e.g., queue status, branch load), and any active consent or policy constraints. This strategy, analogous to retrieval-augmented generation (RAG) \cite{lewis2020rag}, but drawing from live sensor and enterprise context rather than a static document corpus, allows the model to produce responses that are grounded not only in what the user has said, but also in what the system knows about the broader situation.

For example, if a user approaches a kiosk and states that they need help with a transfer, the engine enriches the prompt with the user's authentication status, recent transaction patterns (if consented), and current branch queue information. The resulting response may acknowledge the user's intent, confirm their identity, and proactively offer a streamlined path, all in a single turn, rather than requiring multiple rounds of clarification. This approach reduces conversational overhead and demonstrates the value of ambient context for improving dialogue efficiency \cite{ni2023dialogue, lewis2020rag}.

\subsubsection{Short-Term and Long-Term Memory}

Memory is integral to conversational continuity and personalization. The system maintains a two-tier memory architecture: short-term memory for the ongoing session and long-term memory persisted across sessions.

Short-term memory comprises the running dialogue history and any contextual facts gathered during the current interaction. It enables the agent to maintain coherence within a conversation, track the user's evolving intent, and avoid redundant prompts. The dialogue history is maintained as a structured sequence of turns with role labels (user, assistant) and associated metadata, including timestamps and the contextual signals that were active at each turn.

Long-term memory extends the agent's awareness across sessions. A persistent document store, TinyDB, in the current prototype, with PostgreSQL as scalable production alternatives, records structured summaries of past interactions, user-expressed preferences (e.g., preferred language, communication formality), and salient facts tagged during prior encounters. When composing a response, the engine retrieves relevant long-term memory entries through query-driven lookup and injects them into the prompt context. This mechanism enables behaviors such as referencing a discussion from a previous visit (\textit{``Last time we spoke, you mentioned interest in a savings plan, would you like to continue that conversation?''}) or adapting the agent's tone based on known preferences. Prior research on long-term memory for AI agents highlights that such continuity enhances user trust and perceived competence \cite{zhang2025rag2memory, packer2024memgpt}.

To address privacy and data retention concerns, the system applies purpose-scoped retention policies. Sensitive details are not persisted unless explicitly required, and archived records are subject to anonymization or deletion in accordance with the \emph{temporal horizon and retention} axis of the design space. The system also supports a transparency mechanism through which users can query what the agent remembers about them and request corrections or deletions, aligning with the principle that personalization should remain auditable and user-controlled.

\subsubsection{Proactive Conversational Behaviors}

Beyond responding to user utterances, the conversational layer supports agent-initiated behaviors driven by contextual triggers. These proactive behaviors implement the \emph{initiative and timing} axis of the design space and operationalize anticipatory interaction (R2).

Proactive triggers originate from the sensing layers and are evaluated by the ambient intelligence engine against configurable initiative thresholds. Examples include temporal cues (e.g., the user has been silent for longer than a configurable threshold, $>$2\,s in our prototype, after a system prompt), behavioral cues (e.g., the visual sensing layer detects that the user appears confused or disengaged), and situational cues (e.g., the user's queue number is about to be called, or a previously scheduled appointment time is approaching). When a trigger fires, the engine generates an appropriate prompt offer calibrated to be helpful without being intrusive. The initiative level follows a graduated scale: the agent may remain silent when confidence is low, offer a gentle hint when moderately confident, or provide a specific suggestion when the situational evidence is strong.

This graduated approach is informed by research on mixed-initiative interaction, which shows that users respond favorably to proactive assistance when it is contextually justified and appropriately timed, but negatively to interruptions that are premature or irrelevant \cite{horvitz1999mixed, deng2025proactive}. In early internal evaluations, we observed that context-specific proactive prompts (e.g., \textit{``It looks like you may be waiting for a consultation, would you like me to check on the status?''}) were received more positively than generic offers of help, reinforcing the importance of grounding initiative in the situational state rather than fixed timing rules.

\subsubsection{Transparency and Contextual Disclosure}

A distinctive feature of the conversational layer is the agent's ability to disclose how it uses contextual information. When the digital human draws on ambient signals to inform a response, for instance, referencing appointment data or acknowledging a prior interaction, it can narrate the source of its knowledge (e.g., \textit{``I see from your appointment record that you're here for a mortgage consultation''}). This practice supports the \emph{consent choreography and purpose limitation} constraint by making the system's use of context visible to the user, rather than allowing it to operate as an opaque inference.

Contextual disclosure also extends to the agent's actions. When the digital human performs an operation on the user's behalf (detailed in the following subsection), it verbally confirms the action and its rationale, ensuring that the user remains informed and in control. This narration of context usage and action execution is designed to build trust in an agent that, by nature of its ambient capabilities, has access to information beyond what the user explicitly provides during conversation. The balance between helpfulness and transparency is a recurring design tension in ambient systems \cite{liao2020explainable}, and our approach favors explicit disclosure as the default, with the option for users to adjust the level of explanation through preference settings stored in long-term memory.

\subsection{System Operations}

The utility action layer enables the digital human to move beyond information exchange and perform operations on the user's behalf, such as executing transactions, triggering workflows, generating documents, and coordinating with human staff. This layer transforms the agent from a conversational interface into an active participant in service delivery, capable of completing tasks end-to-end when authorized to do so. The design of this layer is governed by the \emph{actuation scope and safeguards} and \emph{human handoff and organizational coupling} axes of the design space.

\subsubsection{Agentic Task Execution}

The digital human's ability to take action is implemented through an agentic orchestration layer that exposes a set of backend operations as callable functions. The ambient intelligence engine evaluates the conversational and situational context to determine when an action is appropriate, and the LLM can invoke these functions through structured tool-use mechanisms integrated into the generation pipeline \cite{schick2023toolformer, qin2023tool}. Available operations include querying account information, submitting service requests (e.g., ordering a replacement card), scheduling appointments, generating summary documents, and initiating transactional workflows such as fund transfers.

Each operation is defined with a typed JSON Schema specifying its required parameters, preconditions, and expected effects. At generation time, the model can emit a structured tool-call following the function-calling convention supported by OpenAI-compatible APIs and open-weight models such as Llama~3 \cite{patil2023gorilla}. When the agent's reasoning process identifies an applicable action, for instance, if a user states \textit{``I need to block my lost card''}, the engine maps the recognized intent to the corresponding function, validates the required parameters against the current context, and executes the operation through a secure API call to the enterprise backend. The result is then folded back into the conversation: the agent confirms the outcome to the user and updates the session state accordingly. This architecture supports multi-step workflows in which the agent chains several operations in sequence, guided by the evolving dialogue and context. The orchestration design draws on recent work in agentic LLM frameworks such as AutoGen \cite{nine_sixteen_seventeen_wu2023autogenenablingnextgenllm} and ReAct \cite{yao2023react}, which demonstrate that structured tool use and iterative reasoning enable language models to handle complex, multi-turn task completion.

\subsubsection{Actuation Scope and Safeguards}

Not all actions carry equal risk, and the system enforces a graduated model of actuation scope that ties the agent's autonomy to the sensitivity of each operation. We distinguish three tiers of risk:

\begin{itemize}
    \item \textbf{Low-risk actions} include information retrieval, wayfinding guidance, and general recommendations. The agent may perform these autonomously based on situational context, without requiring explicit user approval for each individual action.
    \item \textbf{Medium-risk actions} include form prefilling, document generation, and service request submission. These actions are performed with user confirmation, and the agent presents a summary of the intended operation and proceeds only upon explicit approval.
    \item \textbf{High-risk actions} include financial transactions, account modifications, and operations with regulatory implications. These require both explicit user confirmation and, in some cases, dual control with a human staff member. The system may also require step-up re-authentication (e.g., biometric confirmation on the user's personal device) before executing high-risk operations.
\end{itemize}

This tiered model operationalizes the \emph{actuation scope and safeguards} axis and ensures that the agent's autonomy scales with the consequences of its actions. All executed operations are logged with timestamps, contextual metadata, and the user consent state at the time of execution, creating an auditable record that supports governance and accountability requirements.

The agent is also constrained to a whitelisted set of functions with validated argument schemas. It cannot invoke arbitrary operations or access systems beyond its authorized scope. These constraints are enforced at the engine level, independent of the language model, ensuring that even unexpected model outputs cannot result in unauthorized actions.

\subsubsection{Human Handoff and Organizational Coupling}

The digital human does not operate in isolation; it functions within an organizational context that includes human staff, service protocols, and escalation procedures. The system supports structured handoff mechanisms through which the agent can transfer a user to a human colleague when the situation exceeds the agent's authorized scope, when the user explicitly requests human assistance, or when the agent's confidence in its situational assessment falls below a threshold.

Handoff is implemented as a coordinated transition rather than an abrupt disconnection. When escalation is triggered, the agent notifies available staff through the branch management system, transmits a summary of the interaction context (including the user's stated intent, completed steps, and any relevant enterprise data), and informs the user of the transfer. This ensures that the receiving staff member can continue the service encounter without requiring the user to repeat information, which aligns with cross-channel continuity (R4) and reduces the friction commonly associated with agent-to-human transitions \cite{liu2021handoff}.

Staff availability and role assignment are obtained from the enterprise infrastructural layer: the system knows which staff members are present, their specializations, and their current workload. This information enables intelligent routing, for example, directing a mortgage inquiry to a specialist rather than a general teller. The agent may also coordinate with staff proactively (R2), such as notifying a consultant that their next appointment has arrived and providing a brief context summary before the face-to-face interaction begins.

\subsubsection{Proactive Safety Interventions}

In domains where user safety and financial security are paramount, the utility action layer includes a specialized capability for proactive risk intervention. This mechanism monitors transactional signals and conversational cues for patterns indicative of fraud, social engineering, or other threats. When the enterprise risk engine flags an elevated risk score for a pending operation, or when conversational indicators match known scam scenarios (e.g., urgency language, unusual withdrawal requests, references to external pressure), the digital human intervenes before the action is executed.

The intervention follows the same graduated initiative model used elsewhere in the system. At lower risk levels, the agent may ask a clarifying question to verify intent (\textit{``This transaction is larger than your typical activity,can you confirm this is intentional?''}). At higher risk levels, it may recommend involving a human staff member or temporarily suspend the operation pending additional verification. In all cases, the agent explains the reason for the intervention, maintaining the transparency principle that governs the conversational layer. This proactive safety role illustrates how the ambient digital human can serve as both a helper and a guardian, leveraging contextual awareness not only to streamline service tasks but also to protect users from potential harm \cite{ali2022fraud}.

%% file: sections/4.5_case_study_implementation.tex
\section{Prototype Implementation: A Retail Banking Case Study}
\label{sec:casestudy}

This section provides an overview of a prototype architecture for a digital human assistant that could be deployed in a retail banking customer support environment. The prototype operates in two modalities, an in-branch kiosk where customers interact face-to-face with the agent on a large display, and a remote video-call interface where the digital human appears as a live participant alongside a co-browsing panel. In both settings, the agent draws on the full sensing and actuation stack described earlier, that it perceives the customer through voice and visual channels, retrieves enterprise context (appointment records, account information, transaction history), generates context-enriched dialogue via an LLM backbone, and executes service operations on the customer's behalf when authorized. 

\subsection{Scenario Walkthrough}

In the \emph{in-branch} modality, the ambient intelligence layer is continuously active: overhead depth cameras monitor lobby occupancy, BLE beacons track the spatial distribution of customers, and ambient noise levels are sampled to calibrate speech volume and turn-taking thresholds. When a customer enters the branch and approaches a kiosk, the visual sensing pipeline detects their presence (R1). The customer scans a QR code from their mobile banking app, triggering authentication and a lookup against the branch appointment system that retrieves their name, and appointment details. The ambient intelligence engine simultaneously ingests the environmental state (current occupancy, queue length, advisor availability, and noise level) and synthesizes this into the agent's greeting: the digital human addresses the customer by name, acknowledges their scheduled visit purpose, notes that their advisor will be free in approximately five minutes, and offers to begin a preliminary financial review in the meantime. This behavior illustrates how physical environment awareness (R1) feeds directly into proactive, anticipatory assistance (R2).

When the customer asks about their loan balance and payment history, the agent queries the enterprise CRM after confirming data-access consent. Because occupancy sensors indicate other customers nearby, the agent routes sensitive account details to the customer's personal device via a secure deep link (R4) rather than displaying them on the shared kiosk screen, and lowers its voice volume as an additional privacy adaptation. When the advisor becomes available, signaled by a branch management status update correlated with BLE badge tracking, the system performs a structured handoff, transmitting the conversation transcript, retrieved data, and the customer's stated goals to the advisor's terminal so that the customer need not repeat any information.

In the \emph{remote-call} modality, the digital human appears as a video participant in a browser-based interface. Although in-branch sensors are unavailable, the agent still draws on ambient signals from the digital channel: device metadata (time zone, network quality), recent transaction activity context (e.g., the customer requested a quote for mortgage rate before initiating the call), and enterprise data (open service requests, time since last branch visit). The customer sees the avatar alongside a co-browsing panel that the agent can populate with account dashboards, product comparisons, or document previews, narrating each element as it highlights them on screen (e.g., \textit{``Here is your current outstanding balance, and below you can see your recent payment schedule''}). This transforms the interaction into a visually guided consultation. PII rendering within the co-browsing panel is gated by the same consent and policy layer described in the system architecture (R5, transparency), and the audio pipeline continuously monitors paralinguistic cues; hesitation markers or prolonged silence after a complex explanation may prompt the agent to revisit the topic or simplify its presentation.

\subsection{Human-Like Avatar Rendering}

The visual embodiment of the digital human is a critical determinant of user trust, engagement, and perceived social presence. Research on believable and anthropomorphic agent design \cite{seymour2018actors} demonstrates that avatar realism significantly influences user willingness to interact, particularly in high-stakes service domains where credibility and empathy are essential. In our retail banking prototype, the rendering subsystem produces a realistic, expressive avatar with synchronized speech, facial animation, and gestural behavior driven in real time by the conversational layer. The architecture treats the rendering engine as a pluggable component: Unreal Engine~5 with MetaHuman \cite{epic2021ue5} provides the highest visual fidelity through strand-based hair and subsurface skin scattering; Unity \cite{unity2021hdrp} (via HDRP or URP) offers cross-platform portability to kiosks and mobile devices with lower GPU requirements; and NVIDIA ACE \cite{nvidia2023ace} supplies an engine-agnostic animation backend whose Audio2Face module generates blend-shape weights from speech audio via neural networks. In our preferred configuration, NVIDIA ACE drives the animation pipeline while Unreal MetaHuman handles rendering, and the conversational layer's emotion and discourse annotations are mapped through a behavior controller to facial expressions, gaze shifts, head movement, and idle micro-animations that maintain the impression of a living social presence \cite{bickmore2005social}.

\begin{figure}[t]
  \centering
  \includegraphics[width=0.9\linewidth]{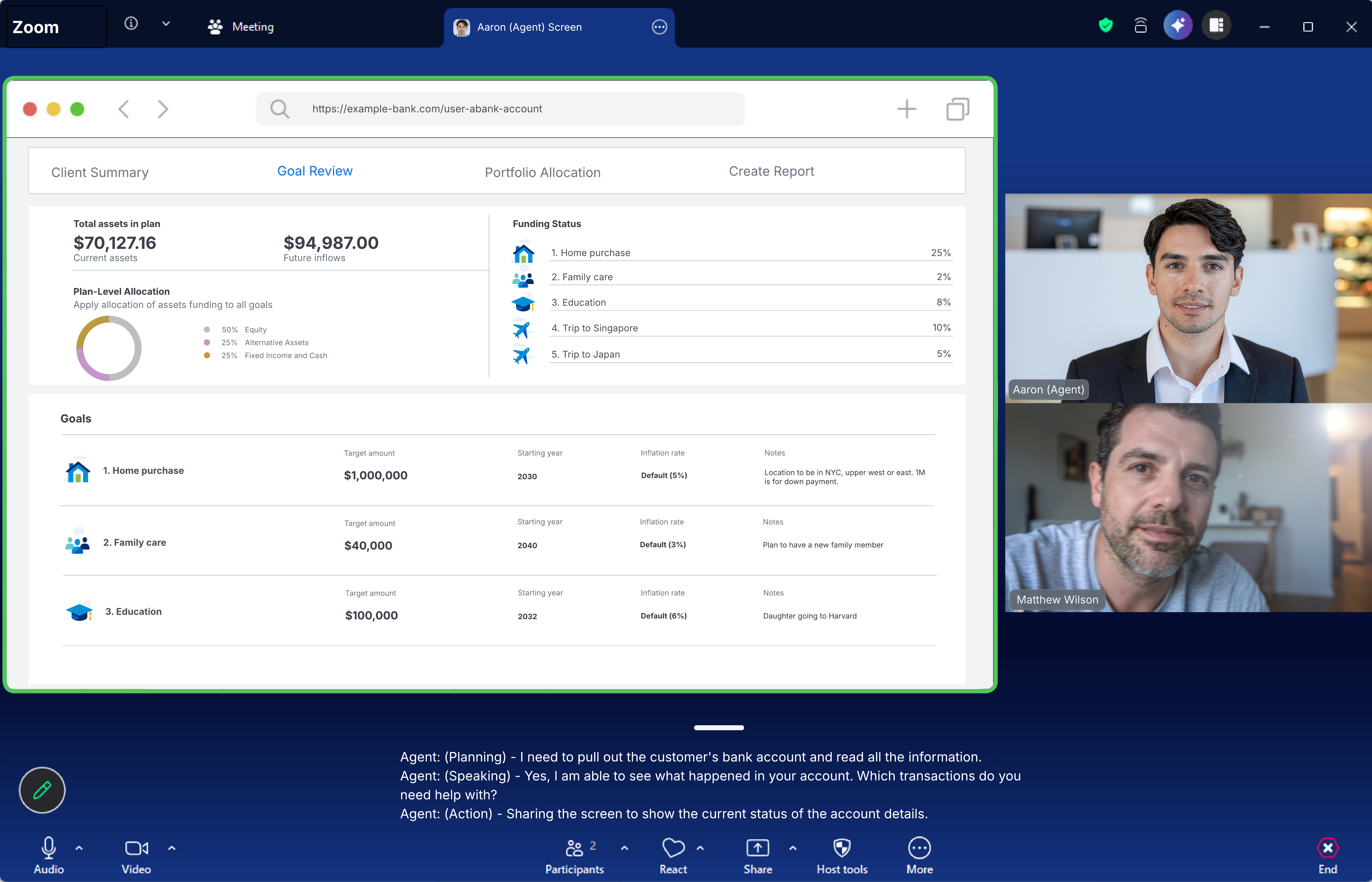}
  \caption{Remote video-call example concept interface for a guided financial review. The digital human is shown in the call tile (top right) alongside a co-browsing panel (left) with a customer account overview.}
  \label{fig:demo_remote}
\end{figure}

\subsection{Frontend Client Interface}

The frontend serves as the customer's primary interaction surface, integrating the avatar video feed, a conversational transcript, and an optional co-browsing view. In the kiosk deployment, the frontend is a full-screen web application running in a locked-down Chromium instance. The avatar occupies the central viewport as a live video stream received via WebRTC from the cloud rendering backend. A chat transcript panel flanks the avatar view, displaying the running dialogue with timestamps. Voice is the primary input channel, captured through an embedded microphone array and streamed to the audio processing service, with an optional on-screen text input for accessibility or noisy environments.

In the remote-call variant, the layout resembles a video conferencing interface. The digital human's video stream occupies one tile, while a second tile displays the agent's co-browsing output. Importantly, the co-browsing model is agent-controlled rather than client-controlled: the digital human operates its own headless browser instance on the server side, navigating enterprise web applications, account dashboards, and product pages as a human advisor would on their own workstation. The resulting browser viewport is captured and streamed to the customer's frontend as a second video or screen-share feed, so the customer observes the agent's navigation in real time without the agent ever accessing or controlling the customer's local browser. When the agent needs to walk the customer through their financial information, it opens the relevant pages in its server-side browser, scrolls to the pertinent sections, and narrates what is being displayed. This mirrors the experience of a human advisor sharing their screen during a consultation. The server-side browser session is ephemeral: it is instantiated per interaction, runs within the customer's authenticated scope (using delegated tokens from the consent and policy layer), and is terminated when the session ends, ensuring that no PII persists beyond the active interaction.

On mobile devices, the layout collapses to a single-column view with the avatar stream above the transcript and the co-browsing feed accessible as a slide-over sheet. Across all form factors, the frontend maintains a persistent WebSocket connection to the ambient intelligence engine for real-time state synchronization, ensuring that proactive prompts and action confirmations appear immediately. This thin-client architecture requires no GPU and minimal compute on the customer's device, as all rendering, inference, and browser automation are performed server-side.

\subsection{Cloud Deployment Architecture}

Because real-time photorealistic rendering is computationally intensive, the prototype adopts a cloud-rendered, stream-to-client architecture rather than requiring dedicated GPU hardware at each service location. Each rendering instance runs as a Docker container that encapsulates the rendering engine (Unreal Pixel Streaming or Unity WebRTC), the animation controller, and the NVIDIA ACE microservices, deployed on Amazon Web Services (AWS) using Elastic Kubernetes Service (EKS) or Elastic Container Service (ECS) with GPU-backed instances (e.g., NVIDIA A10G or A100). EKS provides fine-grained GPU-aware scheduling via the NVIDIA device plugin for Kubernetes and integrates with a broader microservices mesh for complex multi-location deployments; ECS with Fargate reduces operational overhead for simpler configurations where AWS-native task definitions suffice.

Horizontal scaling is managed through auto-scaling policies tied to GPU utilization and active session count. When concurrent rendering sessions exceed the current instance pool's capacity, the auto-scaler provisions additional GPU nodes (in EKS) or tasks (in ECS) to maintain target latency and frame-rate thresholds; during off-peak periods, it scales down to minimize cost. Each rendering instance streams H.264 or AV1 encoded video to the client over WebRTC with adaptive bitrate encoding to accommodate varying network conditions. The non-rendering components (the ambient intelligence engine, audio processor, memory store, and enterprise API adapters) run as separate containerized services on standard (non-GPU) compute, communicating with the rendering tier through internal service mesh endpoints. This separation allows each tier to scale independently: the rendering tier scales with the number of active visual sessions, while the intelligence tier scales with conversational throughput. The entire stack is deployed through a CI/CD pipeline, enabling centralized updates to avatar models, animation assets, frontend builds, and engine logic across all service locations without on-site hardware changes.

%% file: sections/5_Discussion.tex
\section{Discussion}
\label{sec:discussion}

The system architecture presented in the preceding sections demonstrates that the core building blocks for ambient digital humans (multimodal sensing, context-enriched dialogue generation, agentic task execution, and privacy-aware data orchestration) are technically realizable with current methods. Several converging trends reinforce this assessment. Large language models have progressed from text completion engines to multimodal reasoning agents capable of processing interleaved text, image, and audio inputs and of invoking external tools autonomously \cite{openai2023gpt4, grattafiori2024llama3, anthropic2024claude, geminiteam2023gemini, qin2023tool}. Multimodal sensing hardware continues to shrink in cost and power, making deployments of depth cameras, BLE/UWB beacons, and edge-AI accelerators practical in everyday service spaces \cite{gill2024edge}. Concurrently, open interoperability standards such as MCP and Agent-to-Agent protocol are emerging to connect AI agents with data sources, tools, and one another through standardized interfaces. At the same time, organizations in finance, health care, and retail are investing in digital transformation programs that expose enterprise data through standardized APIs, creating the data substrate that ambient agents require \cite{zachariadis2017api}.

Yet a significant gap remains between demonstrating feasibility in controlled prototypes and achieving robust, trustworthy operation in the wild. The challenges are not merely incremental engineering tasks; they touch fundamental open questions in sensing, human--AI interaction, personalization, and governance. In this section we examine five such challenges, situating each within the current state of the art while identifying the specific limitations that ambient digital humans expose and the research directions most likely to address them. Together, these discussions delineate the frontier that must be advanced before the vision articulated in this paper can be fully realized.

\subsection{Robust Multi-Sensor Fusion Under Real-World Variability}

The ambient digital human's situational model (R1) depends on the continuous fusion of signals from heterogeneous sources (microphones, cameras, indoor-positioning infrastructure, device telemetry, and enterprise APIs), each with different noise profiles, update rates, and failure modes. State-of-the-art multi-sensor fusion techniques range from classical Bayesian filters (e.g., extended Kalman filters for localization) to learned fusion models such as multi-modal bottleneck transformers that jointly attend over heterogeneous feature streams through learned attention bottleneck tokens \cite{nagrani2021attention}. In our system, context signals are normalized into a shared JSON representation and combined at the decision layer; however, this late-fusion approach offers limited ability to reason about inter-signal consistency. When a sensor drops out (for example, when a camera feed is temporarily occluded or a BLE beacon becomes unreliable due to interference), the system currently falls back to the remaining signals without a principled estimate of how much the overall confidence should degrade. Similarly, conflicting signals (e.g., the visual pipeline places a user at a counter while BLE localization places them in the waiting area) are resolved by heuristic priority rules rather than by a calibrated uncertainty model.

A more robust architecture would incorporate explicit uncertainty propagation across the fusion pipeline, so that downstream components (dialogue generation, proactive triggers, action gating) can condition their behavior on calibrated confidence intervals rather than point estimates. Promising approaches include \emph{conformal prediction} applied to multi-view sensor fusion \cite{garciaceja2024conformal, cho2024cocoon}, which provides distribution-free coverage guarantees without strong distributional assumptions; probabilistic graphical models that represent sensor reliability as latent variables \cite{khaleghi2013fusion}; and neural evidential reasoning frameworks that place Dirichlet or Normal-Inverse-Gamma priors over predictions to output epistemic uncertainty alongside point estimates in a single forward pass \cite{sensoy2018evidential, amini2020evidential}. A complementary direction is \emph{adaptive fusion scheduling}, where a lightweight policy network learns per-instance modality weights based on signal quality and cross-modal agreement, replacing fixed late-fusion rules with instance-level optimization \cite{panda2021adamml}. Graceful degradation strategies, in which the agent narrows its initiative scope or verbally acknowledges reduced confidence when key signals are unavailable, would align with the \emph{robustness and fallbacks} axis of our design space. Evaluating such strategies in longitudinal field deployments, where sensor failures are not simulated but genuinely unpredictable, remains an important empirical gap.

\subsection{Agentic Browsing and Open Information Retrieval}

A growing expectation for intelligent agents is the ability to autonomously search and retrieve information from the open web and enterprise knowledge bases on behalf of the user, sometimes called \emph{agentic browsing} or, more broadly, \emph{computer use}. In our architecture, the digital human already executes structured tool calls against well-defined enterprise APIs (Section~\ref{sec:actuation}). Extending this capability to unstructured or semi-structured sources (public web pages, PDF documents, internal knowledge portals) or to graphical user interfaces introduces a qualitatively different set of challenges. Agentic web-browsing benchmarks have proliferated since the initial WebArena \cite{zhou2024webarena} and Mind2Web \cite{deng2023mind2web} efforts. More recent evaluations on realistic full-computer tasks, such as OSWorld \cite{xie2024osworld}, report that frontier vision-language models achieve 12--22\% success rates, while web-specific benchmarks such as WebGames report up to 43\% \cite{thomas2025webgames}, both well below human performance (72--96\%). The BrowserGym ecosystem \cite{browsergym2025} provides a unified gym-like environment for standardized agent evaluation across these benchmarks. Commercial computer-use agents, including Anthropic Computer Use and OpenAI Operator \cite{anthropic2024computeruse, openai2025operator}, have brought these capabilities to production, but failure modes persist: incorrect GUI element selection, navigation loops, inability to recover from unexpected page states, and excessive latency. In retrieval-augmented generation (RAG) pipelines, a related limitation is \emph{retrieval noise}: when the agent queries a vector store or search engine, irrelevant or outdated passages may be returned, and recent benchmarking work demonstrates that even top-ranked but non-relevant documents can substantially degrade answer quality \cite{chen2024benchmarking, cuconasu2024noise}. In regulated domains such as finance, where the agent may need to look up current interest rates, regulatory notices, or product terms, factual accuracy is non-negotiable.

Closing this gap requires advances along multiple axes. One avenue is \emph{structured action grounding}: rather than treating web pages as pixel or DOM-level environments, recent work proposes extracting semantic action schemas from web interfaces, analogous to the typed JSON Schema tool definitions used in our system, so that the agent can plan at a higher level of abstraction \cite{boisvert2024workarenaplus}. Another is \emph{proactive self-verification}: emerging approaches such as SmartSnap \cite{cai2025smartsnap} train the agent to proactively collect curated snapshot evidence during GUI task execution using completeness, conciseness, and creativity principles, yielding substantial performance gains over passive post-hoc verification. In practice, however, the most effective near-term strategy may be \emph{hybrid delegation}: letting the agent handle structured API calls autonomously while routing open-ended web searches through a curated, organization-managed knowledge base with editorial oversight. For ambient digital humans in regulated environments, this bounded approach trades some generality for the reliability and auditability that the domain demands.

\subsection{Calibrated Initiative and Proactivity}

The ability to act proactively, offering assistance before the user explicitly asks, is central to the ambient intelligence vision (R2). Yet proactivity is a double-edged capability: well-timed, contextually grounded suggestions are valued, while poorly timed interruptions erode trust and user satisfaction. Our system implements a graduated initiative model in which triggers (silence thresholds, behavioral cues, situational events) are evaluated against configurable confidence levels. This rule-based approach provides transparency and predictability, but it is brittle: fixed thresholds do not account for individual preferences, cultural norms, or task-specific tolerance for interruption. Research on mixed-initiative interaction has long recognized this tension, dating to Horvitz's foundational principles for balancing automated services with direct manipulation \cite{horvitz1999mixed}, yet most deployed systems still rely on manually tuned heuristics. A recent comprehensive survey of proactive conversational AI confirms that no standardized evaluation protocol for agent proactivity has been established \cite{deng2025proactive}. The HCI literature reports that users differ substantially in their preferred level of agent initiative (some welcome frequent suggestions while others find them intrusive), with trust playing a key moderating role across escalating levels of proactive behavior \cite{kraus2021trust}, and that these preferences shift depending on task complexity, emotional state, social context, and individual social-agent orientation \cite{liao2016proactive}.

A more principled approach would be \emph{adaptive initiative modeling}: learning per-user and per-context initiative policies from interaction data. Reinforcement learning from human feedback (RLHF) and contextual bandit formulations have been applied to notification timing, where personalized interruption policies learned from user context significantly improve engagement rates \cite{mehrotra2015notification, tewari2017bandits}. Extending these to the richer state space of an ambient digital human, where the context includes physical signals, dialogue history, and enterprise state, is an open research problem. Notably, while proactive behaviors have appeared in commercial agents (e.g., OpenAI Operator is trained to proactively ask the user to take over for sensitive tasks such as login or payment) and in self-verification systems that seek corroborating evidence during task execution \cite{cai2025smartsnap}, no standardized benchmark or framework exists for evaluating the quality of agent initiative decisions. Any adaptation mechanism must itself be transparent: users should be able to understand why the agent chose to speak or remain silent, and should retain the ability to adjust initiative preferences explicitly. Controlled user studies in realistic service environments, rather than Wizard-of-Oz or crowd-sourced evaluations, are needed to establish empirically grounded baselines for when proactive intervention helps versus hinders the user experience. Our early internal evaluations suggest that grounding prompts in specific situational evidence (e.g., referencing the user's queue status) substantially improves acceptance, but systematic investigation across domains and user populations is still required.

\subsection{Scalable and Privacy-Preserving Personalization}

Long-term memory and personalization (R5) are among the most compelling capabilities of ambient digital humans, and among the most fraught with privacy risk. The tension is structural: richer personalization requires more data, retained for longer, while privacy principles (data minimization, purpose limitation, right to erasure) demand the opposite. Our prototype stores interaction summaries and user preferences in a lightweight document store (TinyDB), with purpose-scoped retention policies and user-accessible memory management. This approach is functional for single-site deployments with modest user populations, but it does not scale gracefully. As the number of users and interaction sessions grows, naive document storage leads to unbounded memory growth; retrieval latency and relevance degrade as the store accumulates heterogeneous records. More fundamentally, storing raw interaction histories, even with retention policies, creates a concentrated data asset that is attractive to attackers and challenging to govern under evolving regulatory frameworks (e.g., the EU AI Act's requirements for high-risk AI systems) \cite{eu2024aiact}. Current state-of-the-art approaches to long-term memory for LLM agents, such as Letta (formerly MemGPT) \cite{packer2024memgpt}, which applies virtual context management inspired by operating-system memory hierarchies, and retrieval-augmented personalization \cite{salemi2024lamp}, focus on memory architecture but give limited attention to the privacy and governance dimensions. Recent benchmarking on very long-term conversational memory further reveals that even long-context LLMs and RAG-augmented systems substantially lag behind human performance in multi-session recall, summarization, and temporal reasoning \cite{maharana2024memory}. The broader research community has recently formalized the umbrella discipline of \emph{context engineering} \cite{mei2025context}, encompassing prompt construction, retrieval strategies, memory management, and tool integration under a unified 166-page survey framework; however, the specific challenges of privacy-preserving long-term context in regulated domains remain underexplored.

Several complementary strategies could address this gap. \emph{Federated and on-device learning} could keep raw interaction data on the user's personal device while sharing only aggregated model updates with the central system, reducing the exposure surface; federated approaches to personalized dialogue generation have demonstrated that persona embeddings can be fine-tuned locally without centralizing private user data \cite{lu2025feddtre, kairouz2021federated}. \emph{Differential privacy} mechanisms can be applied to aggregated behavioral signals (e.g., common navigation patterns, frequently asked topics) so that the system learns population-level trends without retaining individual-level traces; recent work on user-level differential privacy for LLM fine-tuning \cite{charles2024privacy} and on differentially private recommendation \cite{friedman2016privacy} demonstrates practical algorithms for balancing privacy budgets against personalization quality. \emph{Structured memory distillation}, which periodically compresses detailed interaction logs into compact preference profiles and discards the source records, offers a pragmatic middle ground between personalization quality and data minimization. The emerging concept of \emph{stateful agents} \cite{letta2025stateful}, which maintain persistent, evolving memory rather than performing stateless retrieval at each turn, offers an architectural direction that could integrate these strategies: the agent's long-term state can be structured with explicit retention policies, access controls, and distillation schedules built into the memory lifecycle. Standardized context-sharing protocols such as the MCP could further support interoperable, auditable memory exchange across agent boundaries. Ultimately, giving users fine-grained control over what the agent remembers, with intuitive interfaces for reviewing, correcting, and deleting memory entries, would transform personalization from an opaque system process into a collaborative, auditable relationship. Designing and evaluating such interfaces in the context of ambient digital humans is an open HCI research question.

\subsection{Governance, Accountability, and Regulatory Alignment}

As ambient digital humans gain the ability to act autonomously (retrieving sensitive data, executing financial transactions, and coordinating with human staff), questions of governance and accountability become unavoidable. Who is responsible when the agent provides incorrect advice, executes an erroneous transaction, or fails to detect a fraud attempt? Our system addresses governance through several mechanisms: tiered actuation scopes, whitelisted function sets with validated schemas, audit logging of all executed operations, and human-handoff protocols for high-risk scenarios. These measures reflect current best practices for AI governance in regulated industries \cite{jobin2019ethics}, but they are largely static: the risk tiers and function whitelists are defined at design time and do not adapt to evolving threats or regulatory changes. Recent work on adaptive governance argues that generative AI's rapid capability growth demands governance frameworks that co-evolve with the technology rather than relying on rigid one-time provisions \cite{reuel2024governance}. Furthermore, while our audit logs capture \emph{what} the agent did, they offer limited support for explaining \emph{why}. The chain of reasoning from contextual signals through LLM inference to action selection is not fully transparent, a gap that becomes problematic when regulators or compliance officers need to reconstruct a decision after the fact.

The regulatory landscape is no longer prospective. The EU AI Act (Regulation 2024/1689) entered into force in August~2024, with prohibitions on unacceptable-risk AI practices effective since February~2025 and obligations for general-purpose AI (GPAI) models effective since August~2025; the high-risk system transparency and explainability requirements that are most directly relevant to ambient digital humans take effect in August~2026 \cite{eu2024aiact}. The accompanying GPAI Code of Practice \cite{euaioffice2025gpai} introduces additional obligations for foundation models that may underpin systems like ours. Together, these instruments demand capabilities that current prototype architectures do not yet provide, making several research directions urgent rather than speculative. \emph{Runtime explainability}, generating human-readable rationales that reference the specific contextual inputs, policy constraints, and confidence estimates behind each decision, is perhaps the most pressing need. Chain-of-thought prompting and structured reasoning traces \cite{wei2022chain} offer a starting point, but recent analysis cautions that CoT rationales are neither necessary nor sufficient for trustworthy interpretability, as they may not faithfully reflect the model's underlying reasoning process; adapting these techniques to produce rationales that satisfy regulatory standards rather than merely improve accuracy remains an open challenge \cite{chan2025infrastructure}. \emph{Dynamic policy frameworks} that can ingest updated regulatory rules (e.g., new data-sharing restrictions, revised fraud-detection thresholds) and propagate them through the system without requiring redeployment would improve organizational agility. \emph{Consent lifecycle management}, which tracks not just whether consent was obtained but for which purposes, through which channel, and with what expiry, needs to be formalized as a first-class system component rather than handled through ad-hoc checks; existing analyses highlight that the opacity of secondary data processing and inferential analytics fundamentally erode traditional consent models \cite{andreotta2022consent}. Finally, organizational accountability models must clarify the division of responsibility between the agent, the deploying organization, and the technology provider, particularly in cases where the agent operates with partial autonomy. Novelli et~al. propose a seven-feature accountability architecture (context, range, agent, forum, standards, process, and outcomes) that offers a useful starting framework for such clarification \cite{novelli2024accountability}. These questions sit at the intersection of computer science, law, and organizational design, and will require interdisciplinary collaboration to resolve.

\section{Conclusion}

This paper introduced the concept of ambient digital humans, virtual agents that move beyond reactive, dialogue-only interaction by drawing on environmental sensing, cross-device signals, and enterprise data to deliver context-aware, proactive assistance. We presented a conceptual framework that identifies five roles ambient intelligence can play in shaping digital human behavior. Supporting these roles, we defined a layered architecture of ambient context, spanning physical sensing, device and application telemetry, and enterprise infrastructure for information retrieval, alongside conversational, environmental, and utility action channels for system actuation, and organized the resulting design space along input-side and output-side axes that characterize how such systems vary across deployments. We grounded the framework in customer-facing service environments from financial and retail domains, detailing how voice, vision, spatial sensing, device handoff, and enterprise integration are realized in practice. Taken together, our contributions offer researchers and practitioners a structured foundation for designing, building, and evaluating digital humans that respond not only to what users say, but also to where they are, what they are doing, and the broader situational and organizational context in which the interaction takes place.

%% file: appendices/appendix_rescoped.tex

\appendix

\section*{Appendix Overview: A Technical Primer for Digital Human Practitioners}
\label{sec:appendix-overview}

Building an ambient digital human of the kind described in this paper requires integrating technologies from several distinct fields (real-time computer graphics, speech and animation AI, agentic language models, cloud infrastructure, and human perception research) that are rarely covered together in a single treatment. Practitioners entering from any one of these fields typically possess deep expertise in their own domain but limited familiarity with the others: a rendering engineer may have little exposure to agentic AI orchestration, while a conversational-AI researcher may not appreciate why subsurface scattering or strand-based hair physics matter for user trust. Yet, as the main paper argues, the effectiveness of a digital human depends on the \emph{coordinated} quality of all of these components. A photorealistic avatar driven by a shallow chatbot is no more useful than a sophisticated language agent presented through a poorly rendered face.

This appendix is intended to bridge that gap. Rather than a conventional literature review, it is organized as an \textbf{end-to-end technical reference} that traces the full pipeline from visual perception and rendering, through character creation and animation, to speech AI, conversational intelligence, production infrastructure, and emerging generative techniques. Each section is written to be self-contained: a reader can enter at the topic most relevant to their needs without reading the appendix sequentially. Cross-references to the main paper's framework (Sections~\ref{sec:sensing}--\ref{sec:casestudy}) indicate where a given technology is instantiated in our system design, so that the conceptual and the concrete remain connected.

The appendix is structured as follows:
\begin{itemize}
    \item \textbf{Appendix~\ref{sec:appendix-fidelity}} establishes the perceptual and psychological foundations: why visual fidelity is not an aesthetic luxury but a functional prerequisite for trust, emotional communication, and sustained engagement.
    \item \textbf{Appendix~\ref{sec:appendix-rendering}} surveys the real-time rendering techniques (geometry management, global illumination, skin and hair shading) that enable photorealistic digital humans, and compares the principal game engines that implement them.
    \item \textbf{Appendix~\ref{sec:appendix-creation}} covers the upstream asset-creation pipelines (parametric modeling, photogrammetry, sculpting) and the runtime animation and physics systems that bring characters to life.
    \item \textbf{Appendix~\ref{sec:appendix-speech}} describes the AI-driven speech, facial animation, and emotion pipelines, as well as the delivery architectures (cloud streaming, client-side rendering, asynchronous generation) through which digital humans reach end users.
    \item \textbf{Appendix~\ref{sec:appendix-convintel}} provides supplementary depth on conversational intelligence, agentic reasoning, memory systems, and production serving infrastructure, complementing the system-level treatment in the main paper.
    \item \textbf{Appendix~\ref{sec:appendix-frontiers}} surveys generative and neural techniques (GANs, diffusion models, Gaussian splatting, world models, vision-language-action models) that are reshaping the field, alongside standards and interoperability efforts that will shape its future trajectory.
\end{itemize}

\noindent Together, these sections aim to equip researchers and practitioners with a working understanding of the technological landscape that underpins ambient digital human systems, so that design decisions in any one layer can be made with awareness of their implications for the whole.

\section{Visual Fidelity and Human Perception}
\label{sec:appendix-fidelity}

Research in digital human systems often foregrounds backend capabilities (natural language understanding, speech synthesis, and dialogue management), yet empirical evidence from psychology and human--computer interaction consistently demonstrates that the \emph{visual quality} of a virtual agent is an equally critical determinant of interaction outcomes~\cite{mori1970uncanny,macdorman2006uncanny}. This section summarizes the perceptual and cognitive mechanisms through which visual fidelity shapes user responses, motivating the rendering and character-creation technologies surveyed in the subsequent appendices.

\paragraph{The uncanny valley.}
Mori's hypothesis posits that human affinity toward artificial agents rises with increasing realism until a critical threshold, beyond which subtle imperfections in near-human renderings evoke discomfort and unease~\cite{mori1970uncanny}. Subsequent empirical work has validated this effect across cultures and age groups, identifying texture irregularities, asynchronous facial motion, and abnormal eye behavior as primary triggers~\cite{macdorman2006uncanny,mathur2016navigating}. A practical corollary for digital human design is that moderately realistic agents may elicit \emph{less} favorable responses than either clearly stylized or fully photorealistic ones. Systems aiming for sustained user engagement must therefore either adopt an explicitly non-realistic aesthetic or invest sufficiently in rendering quality to cross the uncanny valley threshold.

\paragraph{Trust and credibility.}
Visual realism significantly influences the perceived trustworthiness of virtual agents. Users rate information delivered by realistic avatars as more credible and are more likely to follow their recommendations~\cite{nowak2004influence,nass1994computers}. In healthcare, patients report higher treatment adherence and satisfaction when interacting with photorealistic medical avatars relative to cartoon-like alternatives~\cite{bickmore2010response}. Analogous effects have been observed in financial-service and legal-advisory contexts, where the stakes of misinformation heighten the importance of perceived authority. These findings suggest that visual fidelity functions as a trust signal, and that deploying an under-rendered agent in high-stakes domains may undermine the effectiveness of even a highly capable conversational backend.

\paragraph{Emotional communication.}
Human social cognition relies heavily on facial micro-expressions, gaze dynamics, and nuanced muscle movements to decode emotional state~\cite{ekman2003emotions}. Low-fidelity digital humans cannot reproduce these cues with sufficient resolution, limiting their capacity for affective engagement. Empirical studies confirm that realistic facial rendering enables users to perceive intended emotions more accurately, building rapport and a sense of social presence~\cite{bailenson2006transformed,slater2006place}. This emotional bandwidth is particularly important in applications such as mental health support, customer service, and education, where empathy and interpersonal connection drive outcomes.

\paragraph{Cognitive load and naturalness.}
When an avatar exhibits natural appearance and behavior, users can draw on innate social-cognitive processes, developed through a lifetime of face-to-face interaction, without conscious adaptation~\cite{garau2003impact}. This naturalness lowers the cognitive overhead of the interaction, enabling longer and more productive sessions while reducing fatigue. By contrast, visually unnatural agents impose an additional interpretive burden as users continuously reconcile the agent's appearance with their expectations, diverting attention from the task at hand.

\paragraph{Implications for system design.}
Overall, these findings establish that frontend visual quality is not an aesthetic luxury but a functional requirement that directly conditions the effectiveness of a digital human system. The most sophisticated language model or emotion-recognition pipeline will under-perform its potential if delivered through a visual representation that triggers uncanny-valley effects or fails to convey emotional nuance. This motivates the detailed treatment in the following appendices of rendering techniques~(\S\ref{sec:appendix-rendering}), character creation and animation~(\S\ref{sec:appendix-creation}), and speech and facial animation pipelines~(\S\ref{sec:appendix-speech}) that enable photorealistic digital human deployment.

\section{Real-Time Rendering for Digital Humans}
\label{sec:appendix-rendering}

Rendering a convincing digital human in real time requires the coordinated application of multiple graphics techniques, each addressing a different perceptual cue that the human visual system uses to assess realism. This section surveys the most critical techniques and identifies how they are realized in contemporary game engines, which have become the de facto platforms for interactive digital human applications~\cite{anderson2020surveygames,akeninemuller2019realtime}.

\subsection{Key Rendering Techniques}

\paragraph{Virtualized geometry and level-of-detail.}
Facial surfaces demand extremely high polygon counts to represent pores, wrinkles, and fine anatomical structure, yet rendering millions of triangles per character is prohibitively expensive without intelligent level-of-detail (LOD) management. Virtualized geometry systems address this by decomposing meshes into hierarchical clusters and streaming only the clusters that are visible at the required resolution, as determined by GPU-driven culling at each frame. A digital human's face can thus exhibit full geometric fidelity at close range while automatically simplifying at distance, without requiring manually authored LOD meshes~\cite{karis2021nanite}. Unreal Engine~5's Nanite system is the most prominent implementation of this approach~\cite{epic2021ue5}; Unity~\cite{unity2021hdrp} and CryEngine~\cite{crytek2021cryengine} rely on more traditional artist-authored LOD pipelines.

\paragraph{Global illumination.}
Indirect illumination (light that bounces between surfaces before reaching the viewer) produces the soft color bleeding, ambient gradients, and environmental integration that distinguish photorealistic characters from flat-lit models. Three principal real-time approaches exist. \emph{Hybrid ray-traced global illumination} combines software tracing through signed distance fields, screen-space tracing, and optional hardware ray tracing, accumulating radiance in a temporal cache for stability; Unreal Engine~5's Lumen is the leading example~\cite{hillaire2021lumen}. \emph{Sparse voxel octree global illumination} (SVOGI) voxelizes geometry into a hierarchical structure and traces cones through voxels to gather multi-bounce lighting, capturing off-screen contributions at manageable memory cost; CryEngine pioneered this technique~\cite{crytek2021cryengine}. \emph{Screen-space global illumination} (SSGI) traces rays in screen space to approximate diffuse inter-reflection from visible surfaces, providing a performant but view-dependent alternative available in Unity HDRP~\cite{unity2021hdrp} and other engines. The choice among these methods involves trade-offs between physical accuracy, performance overhead, and hardware requirements.

\paragraph{Subsurface scattering for skin.}
Human skin is translucent: light penetrates the epidermis and dermis, scatters through blood vessels and subcutaneous tissue, and exits at nearby surface points. Simulating this subsurface scattering (SSS) is essential for avoiding the waxy appearance that characterizes skin rendered with purely opaque surface models. The dominant real-time approach uses screen-space separable convolution, blurring diffuse lighting with Gaussian kernels shaped by material-specific scattering profiles~\cite{jimenez2015separable}. The characteristic warm glow of backlit ears and the soft transmission through thin nasal cartilage emerge naturally from this technique. Unity HDRP implements SSS as a core skin-shading feature~\cite{unity2021hdrp}; CryEngine provides multi-layer skin shaders that model epidermis, dermis, and subcutaneous layers with distinct scattering parameters~\cite{crytek2021cryengine}; Unreal Engine~5 includes subsurface-profile-based scattering in its material model~\cite{epic2021ue5}.

\paragraph{Ambient occlusion.}
Contact shadows in facial creases, around the nose bridge, and within wrinkles provide critical depth cues. Screen-space ambient occlusion (SSAO) estimates local occlusion by sampling the depth buffer in a hemisphere around each surface point, producing subtle darkening in concavities~\cite{mittring2007ssao}. First introduced by Crytek in 2007, SSAO and its successors (HBAO+, GTAO) are now universal features in virtually every real-time renderer.

\paragraph{Volumetric lighting and atmospheric effects.}
Participating media such as fog, haze, and volumetric light shafts are rendered by ray-marching through voxelized volume data, accumulating inscattering and attenuation at each step. For digital humans, volumetric effects provide the atmospheric integration that grounds characters in their environments rather than making them appear composited onto a flat background. CryEngine pioneered real-time volumetric rendering~\cite{crytek2021cryengine}, and comparable systems are available in Unreal Engine~5~\cite{epic2021ue5}, Unity HDRP~\cite{unity2021hdrp}, and Frostbite~\cite{ea2020frostbite}.

\paragraph{Physical light units and camera models.}
Photometric light units (lumens, lux, candelas) and physical camera parameters (ISO, aperture, shutter speed) allow lighting artists to apply real-world measurement knowledge directly to digital human scenes. This physically grounded workflow ensures natural integration with live-action footage or photographic reference. Unity HDRP was among the first real-time engines to adopt physical light units as a core feature~\cite{unity2021hdrp}; Unreal Engine~5 and other modern engines offer equivalent capabilities.

\subsection{Hair and Eye Rendering}

Two elements are particularly important for perceptual realism: hair and eyes.

\paragraph{Strand-based hair.}
Photorealistic hair requires rendering tens of thousands of individual strands, each modeled as a curve with properties including thickness, color, melanin concentration, and curliness. Light transport through hair involves multiple scattering components: direct cuticle reflection~(R), transmission through the fiber~(TT), internal reflection followed by transmission~(TRT), and multiple scattering between neighboring strands~\cite{marschner2003light}. The interplay of these components produces specular highlights, translucent back-lighting, and the characteristic depth that distinguishes real hair from opaque shell-based approximations. Unreal Engine's Groom system, Unity's strand-based hair pipeline, and offline tools such as Houdini's XGen implement this approach~\cite{yucel2019hair}. Physics simulation handles strand dynamics through position-based methods with collision against the head mesh and inter-strand self-collision.

\paragraph{Eye rendering.}
The eye comprises multiple optically distinct structures, each requiring dedicated rendering treatment~\cite{jimenez2012next}. The transparent cornea introduces refraction effects, causing the iris to appear to float beneath the surface with correct parallax as the viewing angle changes. The iris itself exhibits fine radial fibers with pigmentation that scatter and absorb light, occasionally producing caustic patterns as the cornea concentrates incoming illumination. The sclera displays subsurface scattering from underlying blood vessels, and its edges exhibit subtle blue tinting. The lacrimal fluid film creates specular reflections, while the meniscus at the lower lid and the caruncle at the inner corner contribute detail that, though often overlooked, is perceptually significant at close range. Pupil dilation responsive to ambient lighting further enhances believability.

\subsection{The Engine Landscape}

The choice of rendering engine depends on target platform, team expertise, licensing model, and required feature depth. Table~\ref{tab:engines} summarizes the principal engines used for digital human applications and their key capabilities.

\begin{table}[t]
\centering
\small
\begin{tabularx}{\linewidth}{l X}
\toprule
\textbf{Engine} & \textbf{Key Digital Human Capabilities} \\
\midrule
Unreal Engine~5 (Epic Games) & Nanite virtualized geometry, Lumen GI, MetaHuman Creator, strand-based hair (Groom), Chaos physics, Pixel Streaming~\cite{epic2021ue5} \\
Unity HDRP & Physical light units, SSS skin shading, SSGI, strand-based hair, broad cross-platform deployment~\cite{unity2021hdrp} \\
CryEngine (Crytek) & SVOGI, multi-layer skin shaders, SSAO (originator), volumetric rendering, performance-capture integration~\cite{crytek2021cryengine} \\
NVIDIA Omniverse & USD-based collaboration, RTX ray tracing, advanced soft-body and physics simulation~\cite{nvidia2021omniverse} \\
O3DE (Open 3D Foundation) & Open-source, CryEngine-derived, integrated cloud services~\cite{amazon2021o3de} \\
Godot Engine & Open-source, Vulkan-based rendering, accessible entry point~\cite{godot2021} \\
\bottomrule
\end{tabularx}
\caption{Principal game engines and platforms for digital human rendering, with their most relevant capabilities.}
\label{tab:engines}
\end{table}

Among these, Unreal Engine~5 currently offers the most integrated feature set for high-fidelity digital humans, combining virtualized geometry, dynamic global illumination, a parametric character creation pipeline, and built-in cloud streaming. Unity provides broader platform reach and greater scripting flexibility, making it well suited if portability to mobile or WebGL targets is a priority. CryEngine retains strengths in foundational rendering techniques it pioneered, while NVIDIA Omniverse occupies a complementary role as a collaborative simulation platform rather than a standalone game engine. The open-source options (O3DE, Godot) offer cost advantages and extensibility, though their digital-human-specific tooling is less mature.

\section{Character Creation, Animation, and Simulation}
\label{sec:appendix-creation}

Rendering quality determines how a digital human \emph{looks}; character creation, animation, and physics simulation determine how it is \emph{built} and how it \emph{moves}. Stiff motion or anatomically implausible deformation can break the illusion of life even when rendering is photorealistic. This section covers the principal workflows for constructing digital human assets and the runtime systems that animate and physically simulate them.

\subsection{Character Creation Pipelines}

\paragraph{Parametric character systems.}
Parametric approaches represent human faces and bodies as continuous vector spaces learned from large collections of high-resolution 3D scans. A user navigates this space by adjusting sliders that blend between shape and texture exemplars, controlling attributes such as nose width, brow height, and lip fullness, while the underlying representation guarantees that all parameter combinations produce anatomically plausible geometry with consistent mesh topology suitable for animation. Unreal Engine's MetaHuman Creator~\cite{epic2021metahuman} is the most widely adopted system of this kind, producing characters with strand-based hair (50,000--100,000 individual strands), anatomical skeletal rigs with over 500 joints, 18 scalable LOD levels, and 180+ FACS-based blendshapes for facial expression~\cite{seymour2021digital}. A machine-learning pipeline can fit the parametric model to arbitrary 3D scans, enabling conversion of custom faces into fully rigged characters. Reallusion's Character Creator~\cite{reallusion2021cc} integrates similar parametric controls with PBR material workflows and multi-layer skin shaders, while Daz3D's Genesis platform~\cite{daz3d2021} uses a unified mesh topology across body types with extensive community-created morph and texture libraries. These parametric pipelines reduce character creation from months of manual work to hours or minutes.

\paragraph{Photogrammetry and 3D scanning.}
When maximal facial fidelity is required, digital humans often begin with direct capture of a real subject. \emph{Light Stage} systems, pioneered by Debevec at USC ICT, surround the subject with hundreds of individually controllable LEDs in a spherical configuration~\cite{debevec2000acquiring}. By photographing the face under each light individually, the system captures the complete reflectance field (how the face appears under arbitrary illumination), yielding diffuse albedo, specular maps, surface normals, subsurface scattering profiles, and displacement maps capturing pore-level geometry. Conventional photogrammetry reconstructs 3D geometry from multiple photographs via structure-from-motion and multi-view stereo algorithms~\cite{beeler2011highquality}. Dynamic (4D) capture systems from vendors such as 3dMD and Ten24 track mesh vertices through time at 60+ frames per second with sub-millimeter accuracy, capturing not only static appearance but the dynamics of expression, specifically how skin stretches and slides over underlying musculature during speech and emotion.

\paragraph{High-resolution sculpting.}
Micro-detail such as skin pores, wrinkles, moles, and facial asymmetry distinguishes individual identities and is essential for crossing the uncanny valley. Digital sculpting tools such as ZBrush~\cite{pixologic2021} operate at tens of millions of polygons, allowing artists to model this fine detail directly. The sculpted detail is subsequently baked into normal and displacement maps for use in real-time rendering, preserving perceptual fidelity at a fraction of the runtime geometry cost.

\paragraph{Cloth and fabric simulation for assets.}
Realistic clothing contributes substantially to digital human believability. Physics-based cloth simulation models garment pattern pieces as 2D shapes with physical properties (mass, stiffness, friction), stitches them together, and drapes them onto a 3D body through iterative position-based dynamics. The resulting garments exhibit naturalistic folds, wrinkles, and draping behavior that would be impractical to achieve through manual sculpting. Marvelous Designer~\cite{marvelous2021} is the leading authoring tool for this workflow; simulated garments are exported as static or animated meshes for use in game engines.

\subsection{Animation Systems}

\paragraph{Skeletal animation and rigging.}
Digital humans are driven by hierarchical skeletal rigs, which are trees of bones whose transformations propagate from parent to child. The visible mesh is bound to these bones through per-vertex skinning weights that determine how much each bone's transformation influences each vertex. Linear blend skinning (LBS) is universal across engines owing to its GPU efficiency, though it produces artifacts (volume loss, candy-wrapper distortion) at extreme joint angles; high-fidelity characters supplement LBS with corrective blendshapes that activate at specific joint configurations to restore anatomically correct deformation. Facial rigs typically contain hundreds of joints replicating underlying muscle groups, driven by FACS-based blendshape systems that decompose any expression into additive combinations of Action Units~\cite{ekman2003emotions}.

\paragraph{Animation state machines and blend trees.}
Interactive digital humans must transition fluidly between behavioral states (idle, speaking, gesturing, listening, reacting) in response to real-time input. Animation state machines define the legal states and transitions, while blend trees interpolate between multiple animation clips based on continuous parameters. Layered animation enables compositing: a base layer handles body posture or locomotion while additive layers apply facial expressions, hand gestures, and head orientation independently. Unreal Engine's Animation Blueprint system, Unity's Animator Controller, and CryEngine's Mannequin system all provide visual graph editors for designing these state machines, enabling seamless transitions without visible discontinuities.

\paragraph{Inverse kinematics and procedural animation.}
Pre-authored animations cannot anticipate every situation a digital human will encounter. Inverse kinematics (IK) computes joint angles that place an end effector at a specified target, enabling a character to maintain eye contact by adjusting head and neck joints toward an interlocutor, reach for objects, or plant feet on uneven surfaces. Full-body IK solvers (FABRIK, CCD) propagate constraints through the entire chain. Procedural animation extends this concept: algorithms generate motion on the fly from rules and physics rather than pre-recorded data. Breathing cycles, weight-shifting during idle poses, blink patterns, and subtle postural sway are commonly generated procedurally to avoid the repetitive quality of looped clips and to maintain the impression of a living presence.

\paragraph{Motion capture integration.}
The most natural digital human motion derives from human performance. Modern pipelines support body tracking (optical marker-based systems such as Vicon and OptiTrack, or markerless computer-vision approaches), facial performance capture (camera arrays or single-camera solutions such as Apple ARKit with TrueDepth), and hand tracking (data gloves or optical finger-level capture). Raw capture data is cleaned, retargeted to the digital human's specific skeletal proportions, and blended with procedural layers. Real-time motion capture further enables live digital human performances in which a human actor drives the avatar simultaneously, a technique used in virtual production, live events, and interactive service applications.

\subsection{Physics Simulation}

Physics simulation ensures that digital humans interact believably with the world: without it, characters appear to float above surfaces, clothing remains rigid, and objects behave unnaturally.

\paragraph{Rigid body dynamics.}
Rigid body simulation governs interactions between a digital human and discrete objects: pressing buttons, grasping cups, opening doors. Collision detection (GJK for convex shapes, bounding-volume hierarchies for spatial partitioning) determines contact events, and constraint solvers compute response forces. Accurate collision ensures that hands rest naturally on surfaces and fingers close plausibly around grasped objects. Unreal Engine uses the Chaos physics engine; Unity integrates NVIDIA PhysX; CryEngine includes its own physics system with soft-body extensions.

\paragraph{Cloth and hair dynamics.}
At runtime, cloth is simulated as a particle system connected by distance, bending, and shear constraints, solved through position-based dynamics (PBD). The simulation accounts for fabric properties, collision with the character's body mesh, and self-collision. Hair simulation follows a similar paradigm, with each strand modeled as a chain of particles with bending and twist constraints; a subset of guide strands is simulated fully, and surrounding strands are interpolated for performance. Wind forces, head motion, and gravity produce natural secondary motion that adds life to the character. Unreal Engine provides Chaos Cloth and the Groom system for strand-based hair physics; Unity offers cloth components and strand-based hair through its Digital Human package.

\paragraph{Secondary motion and soft-body deformation.}
Partial ragdoll or secondary-motion simulation is applied to accessories, jewelry, and soft tissue to add physicality without full ragdoll takeover. Spring-damper systems attached to bones produce subtle secondary motion that follows primary animation with a natural delay. Advanced systems model volumetric soft-tissue deformation (finger-tip compression on contact, thigh flattening when seated) using finite element methods or corrective blendshapes driven by contact events, adding anatomical realism at moderate computational cost.

\subsection{Spatial Audio}

Spatial audio systems model sound propagation through 3D space, reinforcing the perception that a digital human's voice originates from its visible mouth position. Head-related transfer functions (HRTFs) filter audio to simulate how sound waves diffract around the listener's head and pinnae, providing directional cues. Environment-dependent reverb models reflect the acoustic properties of the virtual space (a large hall versus a small room), while occlusion and obstruction models attenuate sound when intervening geometry lies between the speaker and listener. Distance attenuation follows inverse-square falloff consistent with physical acoustics. Unreal Engine's audio system (including MetaSounds for procedural audio), Unity's spatial audio pipeline (with integrations such as Steam Audio and Resonance Audio), and dedicated middleware (Wwise, FMOD) all support these capabilities. For digital humans deployed in physical environments (e.g., kiosks with directional speakers), spatial audio design must additionally account for the real-world acoustic characteristics of the deployment space.

\section{Speech, Facial Animation, and Integrated Platforms}
\label{sec:appendix-speech}

A digital human's visual embodiment is brought to life through the synchronization of speech, facial animation, and expressive behavior. This section surveys the AI-driven pipelines that produce these behaviors and the integrated platforms that package them for deployment, then outlines the principal architectures through which the resulting audiovisual streams are delivered to end users.

\subsection{Audio-Driven Facial Animation}

Generating convincing facial motion from speech audio is a core requirement for interactive digital humans. Deep learning systems trained on hundreds of hours of facial performance data paired with audio learn the correlation between phonemes and visemes (the visual analogs of speech sounds) as well as the mapping from prosodic features (pitch, energy, rhythm) to broader facial expressions~\cite{cudeiro2019captureface}. State-of-the-art models predict approximately 50 blendshape coefficients per frame at 60\,fps, driving standard FACS-based facial rigs directly. Importantly, these systems capture not merely lip movement but full-face animation: brow raises, eye squints, and the subtle asymmetries that humans exhibit during natural speech.

NVIDIA's Audio2Face, part of the ACE platform~\cite{nvidia2023ace}, is the most widely deployed implementation, providing an engine-agnostic animation backend whose neural network generates blend-shape weights from speech audio with naturalistic coarticulatory motion. Soul Machines' Digital Brain~\cite{soulmachines2023} takes a biologically inspired approach, generating micro-expressions, gaze shifts, and postural adjustments from simulated neural dynamics rather than direct audio-to-blendshape regression, producing emergent animation with an organic quality distinct from purely data-driven methods.

\subsection{Speech Recognition and Synthesis}

Real-time spoken interaction requires both automatic speech recognition (ASR) and text-to-speech synthesis (TTS) with stringent latency constraints.

Modern ASR systems employ conformer-based architectures that combine convolutional layers for local feature extraction with transformer attention for global context, achieving sub-200\,ms streaming latency when optimized for GPU inference. TTS systems typically follow a two-stage design: an acoustic model (e.g., FastPitch) predicts mel-spectrograms from text with explicit pitch and duration control, and a neural vocoder (e.g., HiFi-GAN) synthesizes waveforms at 22\,kHz or higher. Voice cloning from approximately 30 minutes of reference audio enables personalized digital human voices that match a specific identity or brand persona.

NVIDIA Riva~\cite{nvidia2023ace} provides production-grade ASR and TTS optimized via TensorRT for low-latency deployment. Replica Studios~\cite{replica2023} specializes in expressive voice synthesis, using variational autoencoders that disentangle content from style, enabling real-time adjustment of emotion intensity, speaking rate, and pitch variation through continuous control vectors. As discussed in the main paper (Section~\ref{sec:sensing}), the choice between cloud-hosted and on-device ASR involves trade-offs between recognition accuracy and data-residency requirements that are particularly important in regulated service environments.

\subsection{Emotion Perception and Expression}

Emotional intelligence is a distinguishing capability of effective digital humans. On the perception side, computer-vision models analyze user facial expressions in real time, estimating emotional states through Action Unit detection from webcam or kiosk camera feeds. This emotional signal can feed back into the digital human's behavior, enabling empathetic response selection and adaptive conversational strategies.

On the expression side, natural language understanding outputs (sentiment polarity, intent confidence, detected topic sensitivity) are mapped to emotional states that modulate facial animation procedurally. Soul Machines' Digital DNA platform~\cite{soulmachines2023} implements a biologically inspired model in which simulated neurotransmitter dynamics produce emotional states that influence both decision-making and expression, yielding emergent animation rather than scripted reactions. NVIDIA ACE~\cite{nvidia2024acedev} integrates emotion perception through its vision pipeline, feeding detected user affect back into Audio2Face expression generation to create responsive, closed-loop emotional interaction.

\subsection{Neural Video Generation}

An alternative to real-time 3D rendering is neural video synthesis, in which generative models produce photorealistic video of a digital human's face asynchronously. The typical pipeline converts input text to speech audio via neural TTS, predicts facial landmarks or blendshape sequences synchronized to that audio using a transformer-based model, and then renders photorealistic video frames conditioned on these control signals and a reference appearance using a generative adversarial network~\cite{thies2020neuralvoice}. Custom avatars can be created from as little as 2--10 minutes of reference video by extracting identity representations (appearance codes, texture maps, and personalized neural rendering models).

Synthesia~\cite{synthesia2023} is the leading platform in this category, supporting multi-language lip-sync and enterprise-scale rendering. Hour One~\cite{hourone2023} offers similar capabilities with especially low capture requirements, using encoder-decoder networks with perceptual loss functions for identity preservation. Neural video generation is well suited to asynchronous content (training videos, informational messages) where real-time interaction is not required; for interactive applications, the latency inherent in frame-by-frame generation currently favors 3D rendering pipelines.

\subsection{Rapid Avatar Creation from Photographs}

Creating personalized 3D avatars from minimal input lowers the barrier to digital human deployment. The dominant approach uses 3D Morphable Models (3DMMs): statistical models learned from thousands of 3D scans that represent faces as linear combinations of shape and texture principal components~\cite{blanz19993dmm}. Given a single photograph, the pipeline detects facial landmarks via a convolutional network, fits the 3DMM through iterative optimization, projects texture from the photograph, inpaints occluded regions using learned priors, and produces a rigged character with PBR materials. Didimo~\cite{didimo2023} implements this pipeline for applications including VR/AR avatars and virtual try-on. Unreal Engine's mesh-to-MetaHuman pipeline~\cite{seymour2021digital} uses a related machine-learning approach to fit parametric models to arbitrary 3D scans.

\subsection{Deployment and Delivery Architectures}

The computational demands of photorealistic digital human rendering typically exceed the capabilities of consumer devices, requiring a server-assisted delivery model. Three principal architectures are in use:

\paragraph{Server-side rendering with cloud streaming.}
The highest visual fidelity is achieved by rendering on GPU-equipped cloud instances and streaming the resulting video to client browsers via WebRTC. WebRTC establishes encrypted, low-latency media channels using DTLS-SRTP transport with adaptive bitrate encoding; congestion-control algorithms estimate available bandwidth and adjust encoder quality in real time~\cite{webrtc2021}. Unreal Engine's Pixel Streaming~\cite{epic2021pixel} and Unity Render Streaming~\cite{unity2021render} are the primary engine-native implementations: both capture rendered frames, compress them via hardware encoders (NVENC, VCE, Quick Sync) with sub-5\,ms encoding latency, and transmit the encoded stream over WebRTC with input events returned via a DataChannel. This approach enables photorealistic characters on any browser-equipped device (including tablets and low-end laptops) at the cost of streaming infrastructure and network-dependent latency. The cloud deployment architecture, including GPU scheduling, horizontal auto-scaling, and edge placement considerations, is discussed in greater detail in the main paper (Section~\ref{sec:casestudy}).

\paragraph{Client-side rendering.}
In-browser rendering via WebGL or WebGPU trades some visual fidelity for broader device compatibility and eliminates streaming latency. UneeQ~\cite{uneeq2023} uses this approach with a lightweight custom engine optimized for facial animation. Client-side rendering is practical for stylized or moderately realistic avatars but currently cannot match the quality of server-rendered pipelines for photorealistic digital humans, particularly on mobile hardware.

\paragraph{Asynchronous neural video generation.}
Platforms such as Synthesia and Hour One pre-render video on GPU clusters and deliver the result as a standard video file or adaptive stream. This model is appropriate for non-interactive content (e.g., personalized onboarding videos, training materials) where real-time responsiveness is not required, and it avoids the infrastructure complexity of live streaming.

Each delivery model involves distinct trade-offs among visual quality, interaction latency, infrastructure cost, and device compatibility. Production deployments often combine architectures: for example, using server-side rendering for in-branch kiosks where a high-speed network is available, and asynchronous generation for email-based follow-up content.

\section{Conversational Intelligence and Production Systems}
\label{sec:appendix-convintel}

A photorealistic, expressively animated digital human is only as useful as the intelligence that drives its behavior. This section provides supplementary technical background on the conversational AI backbone, agentic capabilities, memory systems, and production infrastructure that together allow a digital human to hold coherent conversations, execute tasks, and operate reliably at scale. Many of these topics are discussed in the context of our specific system design in the main paper (Sections~\ref{sec:sensing}--\ref{sec:casestudy}); the treatment here is broader, surveying the general techniques and architectural patterns available to practitioners.

\subsection{Dialogue Generation and Persona Design}

The personality, knowledge boundaries, and behavioral constraints of a digital human are defined primarily through the system prompt and associated guardrails that shape the underlying language model's behavior.

A well-structured system prompt typically specifies: (1)~an \emph{identity definition}: the agent's name, role, organizational affiliation, and background that establishes personality; (2)~\emph{behavioral guidelines}: communication style, verbosity, handling of sensitive topics, and escalation procedures; (3)~\emph{knowledge boundaries}: explicit statements about what the agent does and does not know, limiting confabulation; (4)~\emph{tool instructions}: descriptions of available functions, invocation criteria, and output interpretation; and (5)~\emph{output format directives}: instructions for producing structured outputs such as emotion tags for the animation system or citation markers for source attribution. System prompts are version-controlled alongside application code and deployed through CI/CD pipelines, enabling systematic iteration and regression testing.

Robust safety mechanisms are essential for customer-facing deployment. Input classifiers screen user messages for harmful or adversarial content before they reach the language model, guarding against prompt injection. Output classifiers check generated responses for hallucinated facts, off-brand statements, or policy violations before the digital human speaks them. Grounding the model's responses in retrieved documents or verified data sources, and having the agent attribute information to those sources, further reduces hallucination risk and builds user trust. When the agent detects that it cannot adequately assist (low confidence, emotional distress, legal or medical sensitivity), it initiates a smooth transition to a human agent with full conversational context preserved.

\subsection{Reasoning and Agentic Behavior}

Modern digital humans increasingly function as \emph{agents} that perceive, reason, plan, and act in a continuous loop rather than processing isolated request--response pairs~\cite{park2023generative}. The agent's operational cycle typically comprises perception (encoding multimodal inputs into structured observations), state update (integrating observations with beliefs, memory, and goals), reasoning (deliberating on an appropriate response), action selection (committing to verbal output, tool invocations, or internal updates), execution, and optional reflection on outcomes. This cycle is described in detail for our system in Section~\ref{sec:actuation}; here we briefly note the general reasoning strategies available.

\emph{Chain-of-thought} (CoT) reasoning decomposes complex queries into intermediate steps before generating a final answer, improving accuracy on multi-step problems and providing interpretable traces for auditing~\cite{wei2022chain}. \emph{ReAct} (Reasoning + Acting) interleaves reasoning traces with tool-use actions: the model reasons about what information it needs, calls a tool to obtain it, reasons about the result, and continues until a final answer can be produced~\cite{yao2023react}. For particularly high-stakes decisions, \emph{reflection} mechanisms allow the agent to critique its own draft response for factual accuracy and tone before presenting it to the user.

\paragraph{Tool use and function calling.}
Digital humans that can interact with external systems become substantially more capable. Function calling enables the language model to emit structured JSON invocations against typed API schemas; a middleware layer executes the call and returns the result for incorporation into the response. Operations may include database queries, transaction execution, document generation, appointment scheduling, and third-party service calls. Complex tasks are decomposed into multi-step workflows with intermediate reasoning. The actuation scope, safeguards, and tiered risk model governing these operations in our system are detailed in Section~\ref{sec:actuation}.

\paragraph{Multi-agent orchestration.}
Complex deployments may distribute work across multiple specialized agents: a lightweight router classifies incoming queries and dispatches them to domain-specific agents (billing, technical support, product recommendation), each fine-tuned for its task. Supervisor patterns delegate sub-tasks to worker agents and synthesize their outputs into a coherent response. Frameworks such as LangGraph, AutoGen~\cite{nine_sixteen_seventeen_wu2023autogenenablingnextgenllm}, and CrewAI provide abstractions for message passing, state management, and fault recovery in these architectures.

\subsection{Memory Systems}

Memory transforms a stateless language model into a persistent digital human that recalls past interactions and adapts over time. The main paper describes our system's two-tier memory architecture (Section~\ref{sec:actuation}); here we outline the broader design space.

\emph{Working memory} corresponds to the language model's context window, that is, the token sequence currently being processed. Practical management strategies include sliding-window truncation, periodic summarization of older turns, and hierarchical context layouts that reserve a persistent header for user profile and task state while allocating the remaining budget to recent dialogue.

\emph{Episodic memory} stores records of specific past interactions as dense vector embeddings in a vector database (Pinecone, Weaviate, Milvus, Qdrant, or Chroma), supporting retrieval-augmented generation (RAG) in which semantically relevant past exchanges are injected into the prompt at query time. Temporal indexing enables recency-weighted and time-aware retrieval, and importance scoring (based on emotional intensity, task relevance, or explicit user signals) prioritizes high-value memories during search.

\emph{Semantic memory} captures accumulated knowledge in structured or semi-structured form: user profiles (preferences, expertise, communication style), knowledge graphs linking entities and relationships extracted from conversation, and skill models that track user proficiency to calibrate explanation depth.

\emph{Procedural memory} encodes learned behavioral patterns: successful response strategies reinforced by positive outcomes, error-avoidance records that prevent repetition of known failure modes, and validated workflow templates for recurring multi-step tasks.

Over extended operation, memory stores require active maintenance: consolidation (merging related episodic records into higher-level semantic knowledge), forgetting (time-decay deprioritization of stale entries), contradiction resolution (detecting and updating conflicting facts), and privacy-compliant deletion in response to user requests.

\subsection{Inference Optimization and Serving}

The production backend for a digital human must serve multiple AI models (language model, ASR, TTS, vision, animation) simultaneously under strict latency constraints, as users expect sub-second response times for natural conversation.

\paragraph{Language model inference.}
Large language models are the most computationally demanding component. Key optimization techniques include: \emph{KV-cache management}: systems such as vLLM's PagedAttention treat the key-value cache as virtual memory, allocating non-contiguous blocks on demand to improve GPU utilization; \emph{quantization}: reducing model weights from FP16 to INT8 or INT4 (via GPTQ, AWQ, or hardware-native FP8 on Hopper GPUs) to lower memory requirements with minimal quality loss; \emph{speculative decoding}: a smaller draft model generates candidate token sequences that the larger target model verifies in a single forward pass, achieving 2--3$\times$ speedup when draft acceptance rates are high; and \emph{continuous batching}: dynamically inserting and retiring requests to maximize GPU throughput. Leading inference frameworks include NVIDIA TensorRT-LLM, vLLM, Hugging Face Text Generation Inference (TGI), and SGLang.

\paragraph{Streaming pipeline.}
Conversational digital humans benefit from a streaming architecture in which each component begins processing before its predecessor finishes: (1)~streaming ASR produces partial transcripts as the user speaks, enabling early intent detection; (2)~streaming LLM generation emits tokens incrementally, forwarded to downstream components immediately; (3)~streaming TTS synthesizes audio from the first sentence while later sentences are still being generated; and (4)~streaming animation computes facial blendshape weights from audio chunks in real time. This pipelined approach can reduce end-to-end latency from several seconds to under one second, approaching the pace of natural human conversation.

\paragraph{Microservice decomposition and scaling.}
Production backends typically decompose each capability (ASR, LLM, TTS, animation, memory, tool execution, session management) into independently deployable microservices communicating via gRPC or message queues. Each service scales according to its own demand profile: the LLM tier scales with conversational throughput on GPU nodes, while session management scales on CPU. Service-mesh infrastructure (Istio, Linkerd) provides distributed tracing, traffic management for canary deployments, and mutual TLS. Circuit breakers prevent cascading failures when downstream services degrade, enabling the digital human to acknowledge reduced capability gracefully. GPU cluster management relies on Kubernetes with GPU-aware scheduling (NVIDIA GPU Operator, Triton Inference Server for multi-model serving), horizontal autoscaling tied to queue depth or latency percentiles, and spot instances for cost-effective non-latency-critical workloads such as memory consolidation and model fine-tuning.

\subsection{Observability and Continuous Improvement}

Maintaining quality in production requires comprehensive monitoring. \emph{Turn-level metrics} (response latency, token count, tool-call frequency, error rate) and \emph{session-level metrics} (task completion, satisfaction, escalation and abandonment rates) are tracked for every interaction. Automated evaluators (LLM-as-judge) assess response quality dimensions (relevance, accuracy, helpfulness, tone) on sampled conversations, calibrated against periodic human evaluation. Distributed tracing (OpenTelemetry) follows each interaction through the full pipeline, enabling identification of latency bottlenecks and cross-service error correlation. A/B testing deploys prompt variants, model versions, or pipeline configurations to different user segments, with statistical tests determining which variant performs better. Conversations rated highly by users or evaluators are curated into fine-tuning datasets, closing the feedback loop between deployment and model improvement.

\subsection{Reinforcement Learning from Human Feedback}

Reinforcement learning from human feedback (RLHF) has become a standard technique for aligning language model behavior with human preferences~\cite{ouyang2022instructgpt}. The process comprises three stages: supervised fine-tuning on demonstrations of desired behavior, reward-model training on human-ranked output pairs, and policy optimization (typically via Proximal Policy Optimization) to maximize predicted reward while constraining divergence from the supervised policy. Applied to digital humans, RLHF enables optimization of not only \emph{what} the agent says but \emph{how} it says it: when to speak versus listen, how to handle interruptions, and how to adapt tone to individual users while preserving core identity. Recent variants, including Direct Preference Optimization (DPO), which eliminates the separate reward model, and Constitutional AI, which uses model self-critique to reduce reliance on human labels, further lower the cost of alignment. For deployed digital humans, implicit feedback signals (conversation duration, return visits, task completion) can supplement explicit ratings, enabling continuous online refinement of conversational policies.

\section{Generative Models and Emerging Frontiers}
\label{sec:appendix-frontiers}

This final appendix section surveys generative and neural techniques that are reshaping how digital humans are synthesized, perceived, and situated within virtual environments, and identifies several emerging research directions that are likely to influence the next generation of digital human systems.

\subsection{Generative Adversarial Networks for Appearance Synthesis}

Generative adversarial networks (GANs) introduced a training paradigm in which a generator network learns to produce synthetic images by competing against a discriminator that attempts to distinguish them from real data~\cite{karras2019stylegan}. The StyleGAN family (StyleGAN, StyleGAN2, StyleGAN3) advanced this paradigm through architectural innovations: a learned intermediate latent space~$\mathcal{W}$ whose directions correspond to semantically meaningful attributes (age, expression, hair color), style injection via adaptive instance normalization, and progressive training from low to high resolution. This enabled generation of photorealistic human faces at 1024$\times$1024 and beyond. The disentangled latent space is particularly useful for digital human applications: it allows controlled manipulation of attributes such as aging, expression transfer, and appearance diversification without retraining. Practical applications include generating diverse training data for face-recognition systems, creating unique digital human identities without requiring human subjects, and performing real-time facial reenactment through latent-space interpolation.

\subsection{Diffusion Models}

Diffusion models have demonstrated image quality and diversity that match or exceed GANs across many generative tasks~\cite{ho2020ddpm}. The approach is organized around two phases: a forward process that progressively corrupts data by adding Gaussian noise over many timesteps until it becomes isotropic noise, and a reverse process in which a neural network (typically a U-Net with attention layers) learns to denoise at each step, conditioned on the timestep and optional control signals. Modern sampling schedulers (DDIM, DPM-Solver, consistency models) reduce the required number of denoising steps from hundreds to tens while preserving quality. For digital humans, diffusion models enable text-to-avatar generation guided by CLIP embeddings, appearance editing through inversion techniques (DDIM inversion, null-text inversion) that map real faces into manipulable latent representations, inpainting for occlusion handling, and super-resolution for enhancing cloud-streamed video on the client side.

\subsection{Neural Radiance Fields and Gaussian Splatting}

\paragraph{Neural radiance fields.}
Neural radiance fields (NeRFs) represent scenes as continuous volumetric functions that map 3D coordinates and viewing directions to color and density, trained through differentiable volume rendering to reproduce input photographs~\cite{mildenhall2020nerf}. Dynamic variants such as NerFACE condition the radiance field on expression parameters, enabling animation of photorealistic faces from limited input views~\cite{gafni2021nerface}. NeRFs offer high-quality novel-view synthesis but incur substantial per-pixel computational cost, limiting their suitability for real-time interactive applications.

\paragraph{Gaussian splatting.}
3D~Gaussian splatting has emerged as a more efficient alternative, representing scenes as collections of 3D~Gaussian primitives (ellipsoids characterized by position, covariance, opacity, and spherical-harmonic color coefficients)~\cite{kerbl2023gaussiansplatting}. During rendering, Gaussians are projected onto the image plane and composited via differentiable alpha blending. The explicit, point-based representation achieves 100+\,fps on consumer GPUs, orders of magnitude faster than NeRF's volumetric ray marching, while preserving sharp detail. Importantly, the explicit primitives can be animated by transforming their positions and shapes according to skeletal poses. Recent extensions to dynamic humans (4D~Gaussian splatting) learn deformation fields that drive Gaussian parameters based on body or facial pose, enabling real-time rendering of animatable digital humans from captured performance data. This combination of quality, speed, and animatability positions Gaussian splatting as a promising bridge between classical rasterization-based engines and fully neural rendering.

\subsection{Neural Rendering}

Hybrid approaches that augment traditional rasterization with learned components are an active area of development~\cite{lombardi2018deepappearance}. \emph{Neural textures} replace explicit RGB texel values with learned feature vectors decoded by a small network during rendering, enabling view-dependent effects such as skin translucency and subtle specularity to be represented compactly. \emph{Neural rendering primitives}, for example Meta's Codec Avatars, which represent faces as dense point clouds whose per-point appearance is predicted by a decoder conditioned on expression codes, replace traditional mesh-based representations entirely. \emph{Learned shaders} trained on ground-truth path-traced renderings can approximate complex light transport (global illumination, multiple scattering in hair) at interactive rates. These hybrid techniques currently offer quality–performance trade-offs complementary to those of pure rasterization and pure neural methods; their maturation is likely to blur the boundary between game-engine rendering and generative synthesis, with significant implications for digital human visual fidelity.

\subsection{Multimodal Foundation Models}

Large multimodal models that jointly process text, image, and audio inputs~\cite{openai2023gpt4v} extend the perceptual capabilities available to digital humans. Vision encoders (CLIP, SigLIP, or vision transformers) embed visual input into representations aligned with the language model's space, enabling the agent to interpret webcam feeds, document images, or screen content. For digital humans, multimodal perception supports visual context understanding (interpreting the user's environment or objects shown on screen), real-time emotion perception from facial expressions, gesture interpretation (pointing, hand signals, body language), and grounded dialogue in which the agent refers naturally to shared visual context. The main paper discusses how these capabilities are integrated into the ambient sensing pipeline (Section~\ref{sec:sensing}); we note here that the rapid progress in multimodal model scale and capability, exemplified by GPT-4, Gemini, and open-weight alternatives, is steadily expanding the sensory bandwidth available to digital human systems.

\subsection{World Models and Simulation}

World models learn internal representations of environments that enable prediction, planning, and counterfactual reasoning, capabilities that extend digital humans beyond purely reactive behavior~\cite{ha2018worldmodels}. A world model typically comprises an encoder that compresses observations into a compact latent space, a dynamics model that predicts how the latent state evolves (potentially conditioned on actions), and a decoder that reconstructs observations for visualization or auxiliary training objectives. The dynamics model is the core contribution: by learning environment physics in latent space, an agent can simulate hypothetical futures without costly real-world execution~\cite{lecun2022path}.

\paragraph{Architectural developments.}
The Dreamer family (v1--v3) uses a recurrent state-space model combining deterministic recurrence with stochastic latent variables, training policies entirely within imagined rollouts and achieving sample-efficient learning across diverse control tasks~\cite{hafner2020dreamer}. IRIS reformulates world modeling as autoregressive sequence prediction over discrete visual tokens, drawing on scaling insights from large language models~\cite{micheli2023iris}. Genie~\cite{bruce2024genie} learns generative interactive environments from unlabeled video, inferring a latent action space that enables controllable simulation without explicit action annotation. Its successor, Genie~2~\cite{deepmind2024genie2}, scales the approach to generate consistent, navigable 3D environments from single image prompts, demonstrating object permanence, viewpoint-dependent occlusion, and interactive elements (doors, buttons, graspable objects) that emerge from training on gameplay video. A further iteration, Genie~3~\cite{deepmind2025genie3}, reported improvements in visual fidelity and interactive frame rates through a hybrid diffusion-autoregressive architecture, though independent evaluation of its generalization and physical accuracy remains limited at the time of writing.

On the industry side, NVIDIA Cosmos~\cite{nvidia2025cosmos} provides a family of diffusion-transformer world models trained on large-scale video corpora depicting physical interactions, targeting robotics and embodied-agent applications. Cosmos uses a neural video tokenizer for efficient temporal compression and accepts diverse conditioning signals (text, images, proprioceptive state). Its emphasis on physical plausibility (accurate action--consequence relationships rather than purely visual appeal) makes it relevant for digital humans that must reason about the physical effects of suggested actions. UniSim~\cite{yang2023unisim} pursues a related goal, learning to simulate the outcomes of actions in realistic environments from diverse real-world video. OpenAI's Sora and related large-scale video generation models have demonstrated that video diffusion at scale implicitly acquires world knowledge (object permanence, scene dynamics), though these systems are primarily optimized for generation quality rather than interactive control~\cite{videoworldsimulators2024}.

\paragraph{Relevance to digital humans.}
For digital human systems, world models open several capabilities. An agent equipped with a learned dynamics model can evaluate candidate actions by imagining their outcomes before committing: for example, simulating a proposed workspace arrangement or previewing the visual effect of a financial decision. World models also enable safe exploration in simulation, training in imagined environments (practicing difficult conversational scenarios or physical tasks), and counterfactual reasoning (``what would have happened if the user had chosen option~B?''). The principal open challenges are temporal consistency over long rollouts, physical accuracy in novel scenarios outside the training distribution, and the computational cost of real-time world-model inference alongside the other components of the digital human pipeline.

\subsection{Vision-Language-Action Models and Embodied AI}

Vision-language-action (VLA) models extend multimodal foundation models to output motor actions, creating end-to-end systems that perceive, reason, and act~\cite{brohan2023rt2}. RT-2 fine-tunes vision-language models on robot trajectory data, tokenizing actions as text strings so that the model's language-modeling capability doubles as a motor-command generator. PaLM-E embeds continuous observations directly into the language model's token space for embodied planning~\cite{driess2023palme}. Open-source VLA models such as OpenVLA~\cite{kim2024openvla} and cross-embodiment datasets such as RT-X~\cite{padalkar2023rtx} are lowering the barrier to developing generalizable embodied agents.

For digital humans, VLA capabilities become relevant as systems acquire physical embodiment through telepresence robots and android platforms~\cite{hanson2021sophia}. The central challenge is \emph{personality continuity}, maintaining consistent identity and conversational behavior regardless of whether the agent is rendered on a screen or controlling a physical body. Safe physical interaction (compliant actuators, force sensing, predictive intent models), energy-efficient on-device inference, and haptic perception add further requirements that remain active research areas. The synthesis of VLA agents with world models, enabling planning through imagination and safe simulated exploration, represents a promising direction for embodied digital humans that must operate in uncontrolled physical environments.

\subsection{Evolutionary and Swarm-Based Optimization}

Optimization methods inspired by biological evolution and collective behavior offer complementary tools for digital human system design, though their applicability is more limited than that of the neural techniques surveyed above.

\paragraph{Genetic algorithms.}
Genetic algorithms (GAs) maintain populations of candidate solutions that evolve over generations through selection, crossover, and mutation~\cite{sims1994evolving}. In the digital human context, GAs can explore the latent spaces of generative models (StyleGAN, diffusion models) to discover appearances satisfying complex multi-attribute criteria that are difficult to specify as differentiable objectives. They can also evolve physics-simulation parameters and procedural-animation rules to produce natural-looking movement without extensive motion capture, or optimize multi-objective trade-offs (task completion, user satisfaction, latency) via Pareto-front methods such as NSGA-II. These applications are best understood as specialized search procedures for high-dimensional, non-differentiable design spaces rather than as general-purpose digital human technologies.

\paragraph{Swarm intelligence.}
Swarm methods, including particle swarm optimization (PSO) and ant colony optimization, derive coordinated collective behavior from simple individual rules~\cite{kennedy1995pso}. Potential applications include decentralized coordination of crowd behavior in virtual environments with many digital human characters, distributed hyperparameter tuning across cloud instances, and dynamic load balancing of rendering and inference resources. In multi-agent scenarios with tens or hundreds of digital humans (e.g., background populations in virtual worlds), swarm-based coordination can produce emergent social dynamics (group formation, information propagation, status hierarchies) with lower computational overhead than centralized simulation. These techniques are not substitutes for the neural and learning-based methods described above but may serve useful roles in specific deployment contexts where decentralized, population-level optimization is required.

\subsection{Standards and Interoperability}

As digital human technology matures, standardization efforts are beginning to address the portability, composability, and governance challenges that arise when multiple vendors, engines, and AI backends must interoperate~\cite{metaverse2022standards}.

\paragraph{Avatar and asset formats.}
VRM provides a standardized format for humanoid avatars, specifying rigging conventions, blendshape naming, and material properties. Universal Scene Description (USD) offers interchange for complex assets including animation and physics data. Ready Player Me provides cross-platform avatar creation using a consistent mesh topology. Adoption of these formats enables organizations to decouple avatar creation from rendering engine and deployment platform, reducing vendor lock-in.

\paragraph{Agent communication protocols.}
As digital humans become autonomous agents acting on behalf of users, standardized protocols for agent-to-agent communication, capability advertisement, and trust establishment become necessary. The Model Context Protocol (MCP) \cite{anthropic2024mcp} and Agent-to-Agent (A2A) protocol \cite{google2025a2a} are early efforts in this direction, providing standardized interfaces through which AI agents connect to data sources, tools, and one another. Analogous to how HTTP standardized web communication, such protocols may eventually enable digital humans from different vendors to coordinate seamlessly in shared service environments.

\paragraph{Expression and behavior standards.}
Standardized representations for facial expressions (extending FACS with digital-human-specific conventions), gesture vocabularies, and emotional-state encodings would enable AI systems to generate animation directives that work across different avatar implementations. While no dominant standard has yet emerged, the increasing fragmentation of the ecosystem is likely to accelerate standardization efforts.

\bigskip
\noindent The convergence of real-time graphics, cloud infrastructure, and generative AI positions digital humans as an increasingly viable interface between humans and intelligent systems. The techniques surveyed in this appendix, from physically based rendering and facial animation to agentic architectures and learned world models, provide the technical substrate on which the ambient digital human systems described in the main paper are built. Continued progress across these domains will yield agents that are not only more capable and naturalistic but also more trustworthy, interoperable, and accessible to a broadening range of applications and organizations.